\documentclass[a4paper,fleqn,usenatbib]{mnras}

\usepackage{newtxtext,newtxmath}
\usepackage[T1]{fontenc}

\usepackage[english]{babel}
\usepackage{graphicx}
\usepackage{fleqn}
\usepackage{amssymb}
\usepackage{amsmath}
\usepackage{booktabs}
\usepackage{hyperref}
\usepackage{comment}
\usepackage{enumitem}
\usepackage{breqn}
\usepackage{enumitem}

\makeatletter
\edef\mg@Greek{\hexnumber@\symlettersA}
\makeatother

\setlist[enumerate]{
  labelsep=8pt,
  labelindent=0.\parindent,
 itemindent=0pt,
  leftmargin=*,
}

\usepackage{color, colortbl}

\newcommand*{\mycdot}{\kern-.2em\cdot\kern-.2em}
\renewcommand{\S}{Section}
\newcommand{\F}{Fig.}

\newcommand{\ve}[1]{\boldsymbol{#1}}
\newcommand{\unit}[1]{\hat{\boldsymbol{#1}}}

\newcommand{\msun}{\mathrm{M}_\odot}

\newcommand{\au}{\,\textsc{au}}

\renewcommand{\j}{\jmath}

\newcommand{\feI}{\ve{f}_{\scriptscriptstyle e}^{(0)}}
\newcommand{\feII}{\ve{f}_{\scriptscriptstyle e}^{(1)}}

\newcommand{\fjI}{\ve{f}_{\scriptscriptstyle \j}^{(0)}}
\newcommand{\fjII}{\ve{f}_{\scriptscriptstyle \j}^{(1)}}

\newcommand{\geI}{\ve{g}_{\scriptscriptstyle e}^{(0)}}
\newcommand{\geII}{\ve{g}_{\scriptscriptstyle e}^{(1)}}
\newcommand{\geIII}{\ve{g}_{\scriptscriptstyle e}^{(2)}}

\newcommand{\gjI}{\ve{g}_{\scriptscriptstyle \j}^{(0)}}
\newcommand{\gjII}{\ve{g}_{\scriptscriptstyle \j}^{(1)}}
\newcommand{\gjIII}{\ve{g}_{\scriptscriptstyle \j}^{(2)}}

\newcommand{\heI}{\ve{h}_{\scriptscriptstyle e}^{(0)}}
\newcommand{\heII}{\ve{h}_{\scriptscriptstyle e}^{(1)}}
\newcommand{\heIII}{\ve{h}_{\scriptscriptstyle e}^{(2)}}
\newcommand{\heIIII}{\ve{h}_{\scriptscriptstyle e}^{(3)}}

\newcommand{\hjI}{\ve{h}_{\scriptscriptstyle \j}^{(0)}}
\newcommand{\hjII}{\ve{h}_{\scriptscriptstyle \j}^{(1)}}
\newcommand{\hjIII}{\ve{h}_{\scriptscriptstyle \j}^{(2)}}
\newcommand{\hjIIII}{\ve{h}_{\scriptscriptstyle \j}^{(3)}}

\newcommand{\epssa}{\epsilon_\mathrm{SA}}
\newcommand{\epssao}{\epsilon_\mathrm{SA,1}}
\newcommand{\epssat}{\epsilon_\mathrm{SA,2}}

\newcommand{\epsoct}{\epsilon_\mathrm{oct}}
\newcommand{\epsopn}{\epsilon_\mathrm{1PN}}

\newcommand{\eper}{E}
\newcommand{\mper}{M}
\newcommand{\qper}{Q}

\newcommand{\bin}{ {\mathrm{bin}}}
\newcommand{\rg}{ r_{\mathrm{g}}}

\newcommand{\Leper}{L}

\newcommand{\Ec}{\mathcal{E}}
\newcommand{\Lc}{\mathcal{L}}

\newcommand{\pI}{Paper I}
\newcommand{\Rt}{\mathcal{R}}
\newcommand{\Do}{ \left \langle \Delta \Rt \right \rangle}
\newcommand{\Dt}{ \left \langle (\Delta \Rt)^2 \right \rangle}

\pubyear{2019}

\usepackage{listings}

\usepackage{color}

\definecolor{dkgreen}{rgb}{0,0.6,0}
\definecolor{gray}{rgb}{0.5,0.5,0.5}
\definecolor{mauve}{rgb}{0.58,0,0.82}

\allowdisplaybreaks

\begin{document}
\onecolumn
\title[Secular encounters]{Analytic computation of the secular effects of encounters on a binary: third-order perturbation, octupole, and post-Newtonian terms; steady-state distribution}

\author[Hamers \& Samsing]{Adrian S. Hamers$^{1}$\thanks{E-mail: hamers@ias.edu} and Johan Samsing$^{2}$\thanks{E-mail: jsamsing@gmail.com} \\
$^{1}$Institute for Advanced Study, School of Natural Sciences, Einstein Drive, Princeton, NJ 08540, USA \\
$^{2}$Department of Astrophysical Sciences, Princeton University, Peyton Hall, 4 Ivy Lane, Princeton, NJ 08544, USA}
\date{Accepted 2019 July 20. Received 2019 July 18; in original form 2019 June 19}

\label{firstpage}
\pagerange{\pageref{firstpage}--\pageref{lastpage}}
\maketitle

\begin{abstract}  
Dense stellar systems such as globular clusters are believed to harbor merging binary black holes (BHs). The evolution of such binaries is driven by interactions with other stars, most notably, binary-single interactions. Traditionally, so-called `strong' interactions are believed to be the driving force in this evolution. However, we recently showed that more distant, i.e., `weak' or `secular' encounters, can have important implications for the properties of merging BH binaries in globular clusters. This motivates more detailed understanding of the effects of secular encounters on a binary. In another previous paper, we analytically calculated expressions for the changes of the eccentricity and angular-momentum vectors taking into account second-order (SO) perturbation theory, and showed that, for highly eccentric binaries, the new expressions give rise to behavior that is not captured by first-order (FO) theory. Here, we extend our previous work to third order (TO) perturbation theory. We also include terms up to and including octupole order. The latter are nonzero for binaries with unequal component masses. In addition, we consider the effects of post-Newtonian terms, and we determine the steady-state distribution due to the cumulative effect of secular encounters by computing the associated angular-momentum diffusion coefficients, and applying the Fokker-Planck equation. Together with our previous work, the results in this paper provide a framework for incorporating the effects of distant encounters on binaries in models of cluster evolution, such as Monte Carlo codes. 
\end{abstract}


\section{Introduction}
\label{sect:introduction}
Gravitational interactions in dense stellar systems such as globular clusters can lead to the formation of a dense core with a number of tight black-hole (BH) binaries (e.g., \citealt{1987degc.book.....S,2000ApJ...528L..17P,2008gady.book.....B,2010MNRAS.402..371B,2013MNRAS.429.2997L,2014MNRAS.444...29L,2015PhRvL.115e1101R,2017MNRAS.464L..36A,2017MNRAS.469.4665P}). The evolution of these BH binaries is of great importance for the understanding of BH mergers in dense stellar systems, which have received much interest in the past years with the direct detection of gravitational waves (e.g., \citealt{2016PhRvL.116x1103A,2016PhRvL.116f1102A,2017PhRvL.118v1101A,2017ApJ...851L..35A,2017PhRvL.119n1101A,2017ApJ...848L..12A}). Traditionally, so-called strong encounters, which are associated with energy exchange between the binary and a passing object (a star, or another compact object) are believed to dominate the binary BH evolution. More distant encounters, with periapsis distance $Q\gtrsim 2a$, where $a$ is the binary semimajor axis, are not believed to play an important role, since energy changes are exponentially suppressed (e.g., \citealt{1975MNRAS.173..729H,2003CeMDA..87..411R}), and the eccentricity changes are typically small \citep{1995ApJ...445L.133R,1996MNRAS.282.1064H,2016ApJ...818...21L,2018MNRAS.476.4139H}. 

However, in a recent paper \citep{2019arXiv190607189S}, we showed that these more distant encounters, i.e., weak encounters of the `secular' kind, are in fact important for some properties of merging BH binaries, such as the ratio of the number of mergers occurring inside the cluster to the number of mergers of BH binaries that are ejected from the cluster. Therefore, it is of importance to understand in more detail the effects of secular encounters on binaries in dense stellar clusters.

A seminal work addressing this topic is \citet{1996MNRAS.282.1064H}, who derived expressions for the scalar eccentricity change to the quadrupole and octupole expansion orders in the ratio $r/R$, where $r$ is the binary separation, and $R$ is the perturber separation. In the secular limit, the binary's mean motion is much faster than the perturber's motion, such that it is justified to average the equations of motion (or, more fundamentally, the Hamiltonian) over the binary, resulting in the so-called `single-averaging' (SA) equations of motion. \citet{1996MNRAS.282.1064H} subsequently computed the scalar eccentricity change $\Delta e$ by integrating the equations of motion over the perturber, assuming either a parabolic or hyperbolic orbit.

In their approach, \citet{1996MNRAS.282.1064H} assumed that the binary's eccentricity and angular-momentum vectors are constant during the perturber's passage. This corresponds to a `first-order' (FO) approach in the parameter $\epssa$, which measures the strength of the perturbation (see equation~\ref{eq:epssa} below). However, as we showed in \citeauthor{2019arXiv190409624H} (\citeyear{2019arXiv190409624H}; hereafter \pI) and in \citet{2019arXiv190607189S}, even if the binary is considered to be `hard' in the orbital energy sense, i.e., its orbital energy is much larger than the typical kinetic energy of stars in the cluster \citep{1975MNRAS.173..729H,1993ApJ...403..256H,1996ApJ...467..359H}, it can be `soft' in the angular-momentum sense if its eccentricity is very high. In the latter case, the binary can still be susceptible to perturbations near its apoapsis (note that the binary orbital speed at apoapsis is proportional to $[(1-e)/(1+e)]^{1/2}$). For such binaries that are hard in the orbital energy sense but soft in the angular-momentum sense, the FO expressions can break down, i.e., the eccentricity change according to the FO theory can be such that the new eccentricity is $e' = e+\Delta e>1$. 

In reality, $\Delta e$ is such that $e'<1$ in the secular limit, and the behaviour can be well described by modified expressions that are second order (SO) in $\epssa$, and which were derived in \pI. This was achieved by developing a Fourier expansion of the equations of motion in order to obtain a description of the instantaneous response of the eccentricity and angular-momentum vectors of the binary to the perturber, and substituting the resulting expressions back into the equations of motion. This approach was used in an earlier work in the context of hierarchical triples (i.e., with a bound third object) by \citet{2016MNRAS.458.3060L}. In hierarchical triples, the instantaneous response of the binary to the perturber's phase can lead to long-term modulations of the secular evolution, including orbital flips (e.g., \citealt{1936MNRAS..97...56B,2004AJ....128.2518C,2005MNRAS.358.1361I,2012arXiv1211.4584K,2012ApJ...757...27A,2013PhRvL.111f1106S,2014MNRAS.439.1079A,2014MNRAS.438..573B,2018MNRAS.481.4602L,2018MNRAS.481.4907G}). 

The expressions in \pI~ were valid to SO in $\epssa$, and, for simplicity, we assumed that the octupole expansion-order terms were zero, which is the case for binaries with equal component masses ($m_1=m_2$), and/or when the perturber is very distant. Here, we extend this previous work, and derive expressions to third order (TO) in $\epssa$, and including all associated octupole-order terms. The TO terms give corrections to the lower-order terms, which can be important if $\epssa$ is large (see \S~\ref{sect:TO:test_ex} below for examples). Due to the excessive number of terms in the analytic expressions for hyperbolic perturbers (see Table~\ref{table:N_terms}), we here restrict to parabolic encounter orbits, in which case the number of terms up to including TO in $\epssa$ is more manageable. Our results are incorporated into a freely-available \textsc{Python} script (see the link in \S~\ref{sect:TO:results}). 

In addition, in this paper we study in more detail the effects of post-Newtonian terms combined with secular encounters (\S~\ref{sect:PN}). Also, in \S~\ref{sect:steady_state}, we study analytically the steady-state binary eccentricity distribution of cumulative secular encounters according to our analytic expressions. This provides insight into the importance of weak encounters for the eccentricity evolution of binaries. Also, we show that the FO terms give a significantly different steady-state distribution compared to when both the FO and SO terms are included.

\section{Third-order perturbation theory including octupole expansion order terms}
\label{sect:TO}
\subsection{Overview}
\label{sect:TO:overview}
As in \pI, we consider a binary (component masses $m_1$ and $m_2$, with $m=m_1+m_2$) with semimajor axis $a$ and eccentricity $e$, perturbed by a third passing body with mass $\mper$ and periapsis distance to the binary center of mass $Q>a$. We describe the binary's orientation using its eccentricity $\ve{e}$ and dimensionless angular-momentum vector $\ve{\j}$, where $\j=\sqrt{1-e^2}$, or, alternatively, using orbital elements (see \pI~for the definitions).

\begin{table}
\begin{tabular}{lcc}
\toprule
Terms of order & \multicolumn{2}{c}{Number of terms in $\Delta e$} \\
& $\eper=1$ & $\eper>1$ \\
\midrule
$\epssa$ (all) & 16 & 60 \\
$\quad \epssa$ & 2 & 8 \\
$\quad \epssa \epsoct$ & 14 & 52 \\
\midrule
$\epssa^2$ (all) & 193 & 55,895\\
$\quad \epssa^2$ & 17 & 1,871 \\
$\quad \epssa^2 \epsoct$ & 60 & 16,035 \\
$\quad \epssa^2 \epsoct^2$ & 116 & 37,989 \\
\midrule
$\epssa^3$ (all) & 1,146 & 2,931,541 \\
$\quad \epssa^3$ & 54 & 38,366 \\
$\quad \epssa^3 \epsoct$ & 175 & 289,496 \\
$\quad \epssa^3 \epsoct^2$ & 311 & 856,072 \\
$\quad \epssa^3 \epsoct^3$ & 606 & 1,747,607 \\
\bottomrule
\end{tabular}
\caption{Number of terms appearing in the analytic expressions for $\Delta e$ formulated using the orbital vectors $\ve{e}$ and $\ve{\j}$, obtained by counting the number of terms after applying the \textsc{Expand} function in \textsc{Mathematica}. For the FO, SO, and TO terms in $\epssa$, we give the total number of terms (indicated with `all') including all octupole-order terms. We also separately state the number of terms associated with different orders of the octupole parameter $\epsoct$. The second and third columns correspond to parabolic ($\eper=1$) and hyperbolic ($\eper>1$) perturber orbits, respectively. }
\label{table:N_terms}
\end{table}

The perturber can be on a parabolic orbit (eccentricity $\eper=1$), or a hyperbolic orbit ($\eper>1$). Although we derived all results for both parabolic and hyperbolic orbits, we do not present results for hyperbolic orbits since the number of terms in this case to high orders of $\epssa$ and/or including octupole-order expansion terms is excessively large. This is illustrated in Table~\ref{table:N_terms}, in which the number of terms appearing in our expressions for the scalar eccentricity change $\Delta e$ is given for the first, second, and third perturbation orders $\epssa$, and for different orders of the expansion octupole parameter $\epsoct$ (see below for the definitions of $\epssa$ and $\epsoct$). The number of terms in the case of hyperbolic orbits is close to 3 million for the $\epssa^3$ term (including all octupole-order terms). In the case of parabolic orbits, the number of terms at order $\epssa^3$ is three orders of magnitudes lower, and still manageable (at least, using computer algebra software), with 1,146 terms.

We consider secular encounters, i.e., encounters for which the perturber angular speed at all times is much lower than the binary mean motion (typically, this implies that the binary is `hard' in the energy sense). The strength of the perturber is characterized by the quantity $\epssa$, which is defined as (cf. \pI)
\begin{align}
\label{eq:epssa}
\epssa \equiv \left [ \frac{\mper^2}{m(m+\mper)} \left ( \frac{a}{\qper} \right )^3 \left(1+\eper \right )^{-3} \right ]^{1/2}.
\end{align}
For equal-mass systems ($m_1=m_2=\mper$) and parabolic orbits, this reduces to $\epssa = (a/Q)^{3/2}/(4\sqrt{3})$.

As in \pI, we expand the Hamiltonian in the ratio of the binary separation, $r$, to the perturber's separation $R$ to the binary's center of mass, and average over the binary's orbital phase. Up to and including third order in $(r/R)$ (i.e., octupole order), the resulting single-averaged (SA) equations of motion read
\begin{subequations}
\label{eq:EOM_SA_gen}
\begin{align}
\nonumber \frac{\mathrm{d}\ve{e}}{\mathrm{d} \theta} &= \epssa (1+\eper \cos \theta) \Biggl \{ -3 \left(\ve{\j}\times \ve{e}\right) - \frac{3}{2} \left (\ve{\j} \cdot \unit{R} \right) \left(\ve{e} \times \unit{R} 
\right )+ \frac{15}{2} \left( \ve{e}\cdot \unit{R} \right ) \left ( \ve{\j} \times \unit{R} \right )\\
& + \epsoct (1+\eper \cos \theta) \frac{15}{16} { \Biggl [ 16 \left(\ve{e} \cdot \unit{R} \right ) \left(\ve{\j}\times \ve{e} \right ) - \left(1-8e^2 \right) \left(\ve{\j} \times \unit{R} \right ) + 10 \left ( \ve{e} \cdot \unit{R} \right ) \left ( \ve{\j} \cdot \unit{R} \right ) \left ( \ve{e} \times \unit{R} \right ) + 5 \left(\ve{\j}\cdot \unit{R} \right )^2 \left ( \ve{\j}\times \unit{R} \right ) - 35 \left ( \ve{e} \cdot \unit{R} \right )^2 \left ( \ve{\j} \times \unit{R} \right ) \Biggl ] \Biggl \} }; \\
\nonumber \frac{\mathrm{d}\ve{\j}}{\mathrm{d} \theta} &= \epssa (1+\eper \cos \theta) \Biggl \{ -\frac{3}{2}  \left ( \ve{\j} \cdot \unit{R} \right ) \left( \ve{\j} \times \unit{R} \right ) + \frac{15}{2} \left ( \ve{e} \cdot \unit{R} \right ) \left ( \ve{e} \times \unit{R} \right ) \\
& +  \epsoct (1+\eper \cos \theta) \frac{15}{16} \Biggl [ - \left(1-8e^2 \right) \left(\ve{e} \times \unit{R} \right ) + 10 \left ( \ve{e} \cdot \unit{R} \right ) \left ( \ve{\j} \cdot \unit{R} \right ) \left ( \ve{\j} \times \unit{R} \right ) + 5 \left(\ve{\j}\cdot \unit{R} \right )^2 \left ( \ve{e}\times \unit{R} \right ) - 35 \left ( \ve{e} \cdot \unit{R} \right )^2 \left ( \ve{e} \times \unit{R} \right ) \Biggl ] \Biggl \}.
\end{align}
\end{subequations}
Here, $\theta$ is the true anomaly of the perturber's orbit; generally, $-\Leper<\theta<\Leper$, where
\begin{align}
\label{eq:theta0_def}
\Leper\equiv \arccos\left ( -\frac{1}{\eper} \right ).
\end{align}
Evidently, $L=\pi$ for parabolic orbits. The octupole-order terms in equation~(\ref{eq:EOM_SA_gen}) are associated with the `octupole parameter' $\epsoct$ (analogous to the octupole parameter in hierarchical triples, e.g., \citealt{2011ApJ...742...94L,2011PhRvL.107r1101K,2013ApJ...779..166T,2014ApJ...791...86L}, or more generally in hierarchical systems, \citealt{2016MNRAS.459.2827H}), which is defined according to
\begin{align}
\label{eq:epsoct}
\epsoct \equiv \frac{|m_1-m_2|}{m_1+m_2} \frac{a}{Q}  \frac{1}{1+\eper}.
\end{align}
Note that the quadrupole-order terms are $\propto \epssa (1+\eper \cos \theta)$, whereas the octupole-order terms are $\propto \epssa \epsoct (1+\eper \cos \theta)^2$. In \pI, we focused on equal-mass binaries ($m_1=m_2$), such that $\epsoct=0$, independent of $a/Q$. Here, we relax this assumption, and derive the expressions including the octupole-order terms.

\subsection{Analytic results}
\label{sect:TO:results}
We follow the same procedure as in \pI~(which was based on \citealt{2016MNRAS.458.3060L}) to derive the changes of the eccentricity and angular-momentum vectors. In summary, we develop the Fourier expansions of the equations of motion, equations~(\ref{eq:EOM_SA_gen}), in terms of $\theta$. Assuming that $\ve{e}$ and $\ve{\j}$ can be written as Fourier expansions of $\theta$, we subsequently derive relations between the Fourier coefficients of $\ve{e}$ and $\ve{\j}$, and the known Fourier coefficients of the equations of motion. We then integrate the equations of motion taking into account the Fourier expansions of $\ve{e}$ and $\ve{\j}$. Physically, this amounts to taking into account the instantaneous response of the binary to the perturber when calculating the net changes to the binary's orbital vectors. For more details on the procedure, we refer to \pI.

We denote the results for the changes of the eccentricity and angular momentum vectors in the following compact form,
\begin{subequations}
\label{eq:Delta_e_Delta_j}
\begin{align}
\Delta \ve{e} &= \epssa \left [ \feI + \epsoct \feII \right ] + \epssa^2 \left [ \geI + \epsoct \geII + \epsoct^2 \geIII \right ] + \epssa^3 \left [ \heI + \epsoct \heII + \epsoct^2 \heIII + \epsoct^3 \heIIII \right ]; \\
\Delta \ve{\j} &= \epssa \left [ \fjI + \epsoct \fjII \right ] + \epssa^2 \left [ \gjI + \epsoct \gjII + \epsoct^2 \gjIII \right ] + \epssa^3 \left [ \hjI + \epsoct \hjII + \epsoct^2 \hjIII + \epsoct^3 \hjIIII \right ].
\end{align}
\end{subequations}
Here, $\ve{f}^{(i)}$, $\ve{g}^{(i)}$, and $\ve{h}^{(i)}$ (with subscripts indicating association with $\Delta \ve{e}$ or $\Delta \ve{\j}$) are vector functions of the initial binary eccentricity and angular-momentum vectors (or, equivalently, the initial binary orbital elements); $i$ in $\ve{f}^{(i)}$ denotes the order of $\epsoct$ with which $\ve{f}^{(i)}$ is associated (similarly for $\ve{g}^{(i)}$ and $\ve{h}^{(i)}$). For small eccentricity changes, the scalar eccentricity change is given by $\Delta e = \unit{e} \cdot \Delta \ve{e}$. Below, we give the explicit expressions for the relevant terms $\unit{e} \cdot \feI$, etc., up to and including TO in $\epssa$, and valid for parabolic orbits (expressed in terms of the orbital vectors). We include the complete octupole expressions for terms FO and SO in $\epssa$, whereas we do not explicitly show the octupole-order terms associated with $\epssa^3$, since the latter expressions are excessively long. The complete expressions for the vector functions appearing in equations~(\ref{eq:Delta_e_Delta_j}) are given explicitly in Appendix~\ref{app:fgh} for the case of parabolic orbits, where we exclude the octupole-order terms at TO in $\epssa$. The complete TO expressions in $\epssa$, i.e., including all octupole-order terms, are implemented in a freely-available \textsc{Python} script\footnote{\href{https://github.com/hamers/flybys}{https://github.com/hamers/flybys}}.

\begin{align}
\nonumber \Delta e &= \frac{15}{2e} \pi  e_z \epssa (e_y \j_x-e_x \j_y) - \frac{15}{32e} \pi  \epssa \epsoct \Biggl [e_x^2 (3 e_y \j_z-73 e_z \j_y)+10 e_x \j_x (7 e_y e_z+\j_y \j_z)+e_y \j_z \left(3 e_y^2-32 e_z^2-15 \j_x^2-5 \j_y^2+4\right) \\
\nonumber &\quad +e_z \j_y\left(-3 e_y^2+32 e_z^2+5 \j_x^2+5 \j_y^2-4\right)\Biggl ] \\
\nonumber &+ \frac{3}{8e} \pi \epssa^2 \Biggl [ e_z^2 \left(-75 \pi  e_x^2+18 \pi  \j_x^2-25 \j_x \j_y+18 \pi  \j_y^2\right)+3 \pi  e_x^2 \left(6 \j_y^2-\j_z^2\right)+e_y \left(e_x \left(-36 \pi  \j_x \j_y-25 \j_y^2+25 \j_z^2\right)+2 e_z \j_z (25 \j_x+6 \pi  \j_y)\right) \\
\nonumber &\quad +2 e_x e_z \j_z (6 \pi  \j_x-25 \j_y)+e_y^2 \left(-3 \pi  \left(25 e_z^2+\j_z^2\right)+18 \pi  \j_x^2+25 \j_x \j_y\right)\Biggl] \\
\nonumber &-\frac{15}{512e} \pi  \epssa^2 \epsoct \Biggl[ e_y \biggl \{ 7 \j_x^2 \left(121 e_x^2+682 e_z^2-\j_y^2-91 \j_z^2-20\right)-6 \j_x \left(662 \pi  e_x^2 \j_y-1281 e_x e_z \j_z+2 \pi  \j_y \left(-16 e_z^2-15 \j_y^2+10 \j_z^2+12\right)\right) \\
\nonumber &\quad +e_x^2 \left(3549 \j_z^2-3066 \j_y^2\right)+2712 \pi  e_x e_z \j_y \j_z+42 e_z^2 \left(73 \j_y^2-80 \j_z^2\right)+49 \j_x^4+180 \pi  \j_x^3 \j_y+21 \left(20-43 \j_y^2\right) \j_z^2\biggl \}+3 e_x^2 e_z \j_z (1100 \pi  \j_x-3759 \j_y) \\
\nonumber &\quad +e_y^2 \left(e_x \left(-12 \pi  \left(730 e_z^2+\j_y^2+15 \j_z^2\right)+1992 \pi  \j_x^2+3773 \j_x \j_y\right)+21 e_z \j_z (28 \pi  \j_x-243 \j_y)\right)+e_x \biggl \{ 12 \pi  \j_x^2 \left(276 e_z^2+5 \left(\j_z^2-3 \j_y^2\right)\right)\\
\nonumber &\quad +7 \j_x \j_y \left(-676 e_z^2+\j_y^2-62 \j_z^2+20\right)+12 \pi  \left(320 e_z^4+4 e_z^2 \left(65 \j_y^2+8 \j_z^2-10\right)-15 \j_y^4+3 \j_y^2 \left(5 \j_z^2+4\right)-4 \j_z^2\right)-49 \j_x^3 \j_y\biggl \} \\
\nonumber &\quad +e_y^3 \left(-707 \j_x^2+12 \pi  \j_x \j_y+2037 \j_z^2\right)+3 e_z \j_z \left(7 \j_y \left(160 e_z^2+51 \j_x^2-20\right)-4 \pi  \j_x \left(96 e_z^2+35 \j_x^2-12\right)-140 \pi  \j_x \j_y^2+301 \j_y^3\right) \\
\nonumber &\quad + e_x^3 \left(-\left(60 \pi  \left(146 e_z^2-33 \j_y^2+3 \j_z^2\right)+847 \j_x \j_y\right)\right) \Biggl] \\
\nonumber &-\frac{225}{8192e} \pi  \epssa^2 \epsoct^2 \Biggl [e_x^4 \left(4 \pi  \left(9 e_y^2+5329 e_z^2+54 \left(\j_z^2-19 \j_y^2\right)\right)+7343 \j_x \j_y\right)+e_x^3 \biggl \{ 10 e_z \j_z (2191 \j_y-1216 \pi  \j_x)-7 e_y \left(1049 \j_x^2-1280 \pi  \j_x \j_y \right. \\
\nonumber &\quad \left. -2045 \j_y^2+612 \j_z^2\right)\biggl \}+e_x^2 \biggl \{ -8 \pi  \j_x^2 \left(602 e_y^2+1342 e_z^2-85 \j_y^2-105 \j_z^2\right)-\j_x \left(2 \j_y \left(6048 e_y^2-4207 e_z^2-1901 \j_z^2+336\right)+14350 e_y e_z \j_z \right. \\
\nonumber &\quad \left. +2237 \j_y^3\right)+8 \pi  \left(9 e_y^4+e_y^2 \left(2573 e_z^2+64 \j_y^2+18 \j_z^2+12\right)-608 e_y e_z \j_y \j_z-2336 e_z^4-2 e_z^2 \left(871 \j_y^2+48 \j_z^2-146\right)+135 \j_y^4-65 \j_y^2 \j_z^2-108 \j_y^2  \right. \\
\nonumber &\quad \left. +12 \j_z^2\right)-281 \j_x^3 \j_y\biggl \}+e_x \biggl \{ e_y^3 \left(-\left(2219 \j_x^2+192 \pi  \j_x \j_y-7679 \j_y^2+2772 \j_z^2\right)\right)+2 e_y^2 e_z \j_z (9359 \j_y-2656 \pi  \j_x)+e_y \left(\j_x^2 \left(-11690 e_z^2+4032 \j_y^2 \right. \right. \\
\nonumber &\quad \left. \left. +722 \j_z^2+672\right)-32 \pi  \j_x \j_y \left(-164 e_z^2+40 \j_y^2+15 \j_z^2-22\right)-2 \j_y^2 \left(2219 e_z^2-3839 \j_z^2+504\right)+2352 \left(8 e_z^2-1\right) \j_z^2+281 \j_x^4-480 \pi  \j_x^3 \j_y \right. \\
\nonumber &\quad \left. +55 \j_y^4\right)+2 e_z \j_z \left(\j_y \left(-30016 e_z^2-8817 \j_x^2+3752\right)+16 \pi  \j_x \left(304 e_z^2+105 \j_x^2-38\right)+2800 \pi  \j_x \j_y^2-7029 \j_y^3\right)\biggl \}+36 \pi  e_y^6-e_y^4 \biggl \{ 4 \pi  \left(183 e_z^2 \right. \\
\nonumber &\quad \left. -10 \j_y^2+18 \j_z^2-24\right)+240 \pi  \j_x^2+7679 \j_x \j_y\biggl \}+2 e_y^3 e_z \j_z (992 \pi  \j_y-4067 \j_x)+e_y^2 \biggl \{ -40 \pi  \j_x^2 \left(2 e_z^2-30 \j_y^2-39 \j_z^2+8\right)-\j_x \j_y \left(3374 e_z^2+55 \j_y^2 \right. \\
\nonumber &\quad \left. +3802 \j_z^2-1008\right)+4 \pi  \left(832 e_z^4+4 e_z^2 \left(123 \j_y^2+48 \j_z^2-58\right)+125 \j_y^4-10 \j_y^2 \left(7 \j_z^2+12\right)-24 \j_z^2+16\right)+300 \pi  \j_x^4-1795 \j_x^3 \j_y\biggl \} \\
\nonumber &\quad +2 e_y e_z \j_z \biggl \{\j_x \left(19264 e_z^2+3945 \j_y^2-2408\right)+16 \pi  \j_y \left(272 e_z^2+55 \j_y^2-34\right)+5517 \j_x^3-240 \pi  \j_x^2 \j_y\biggl \}+4 e_z^2 \biggl \{ 10 \pi  \j_x^2 \left(64 e_z^2-17 \j_y^2-8\right)\\
\nonumber &\quad +3 \j_x \j_y \left(224 e_z^2+191 \j_y^2-28\right)+\pi  \left(1024 e_z^4-128 e_z^2 \left(11 \j_y^2+2\right)-245 \j_y^4+176 \j_y^2+16\right)+75 \pi  \j_x^4+519 \j_x^3 \j_y\biggl \}\Biggl ] \\
\nonumber &+\frac{3}{512e} \pi  \epssa^3 \Biggl [-1200 \pi  e_x^4 \j_z+15 e_x^3 \left(466 e_y \j_z+80 \pi  e_z \j_x+\left(369+384 \pi ^2\right) e_z \j_y\right)-3 e_x^2 \biggl \{ 1000 \pi  e_y^2 \j_z+15 e_y e_z \left(\left(128 \pi ^2-159\right) \j_x+160 \pi  \j_y\right) \\
\nonumber &\quad +2 \j_z \left(20 \pi  \left(160 e_z^2+\j_y^2\right)+120 \pi  \j_x^2+\left(581+192 \pi ^2\right) \j_x \j_y\right)\biggl \}+e_x \biggl \{ 9390 e_y^3 \j_z+15 e_y^2 e_z \left(40 \pi  \j_x+\left(811+384 \pi ^2\right) \j_y\right)+4 e_y \j_z \left(840 e_z^2  \right. \\
\nonumber &\quad \left. +8 \left(36 \pi ^2-7\right) \j_x^2+180 \pi  \j_x \j_y-3 \left(217+96 \pi ^2\right) \j_y^2\right)+e_z \left(-120 e_z^2 \left(80 \pi  \j_x+\left(133-48 \pi ^2\right) \j_y\right)+720 \pi  \j_x^3-313 \j_x^2 \j_y+120 \pi  \j_x \left(9 \j_y^2+32 \j_z^2\right)\right.\\
\nonumber &\quad \left. -927 \j_y^3-616 \j_y \j_z^2+384 \pi ^2 \j_y \j_z^2\right)\biggl \}-1800 \pi  e_y^4 \j_z-15 e_y^3 e_z \left(\left(384 \pi ^2-387\right) \j_x+520 \pi  \j_y\right)-6 e_y^2 \j_z \biggl \{ 20 \pi  \left(80 e_z^2-7 \j_y^2\right)+80 \pi  \j_x^2+\left(371 \right. \\
\nonumber &\quad \left. -192 \pi ^2\right) \j_x \j_y\biggl \}+e_y e_z \biggl \{ -360 \left(21+16 \pi ^2\right) e_z^2 \j_x+87 \j_x^3+720 \pi  \j_x^2 \j_y+285 \j_x \j_y^2-8 \left(7+48 \pi ^2\right) \j_x \j_z^2+120 \pi  \j_y \left(9 \j_y^2+16 \j_z^2\right)\biggl \}\\
\nonumber &\quad +96 e_z^2 \j_x \j_z (20 \pi  \j_x+49 \j_y)\Biggl ] \\
& + \epssa^3 \mathcal{O}(\epsoct) + \mathcal{O} \left (\epssa^4 \right ).
\end{align}

\subsection{Tests and examples}
\label{sect:TO:test_ex}

\begin{figure}
\center
\includegraphics[scale = 0.46, trim = 8mm 0mm 8mm 0mm]{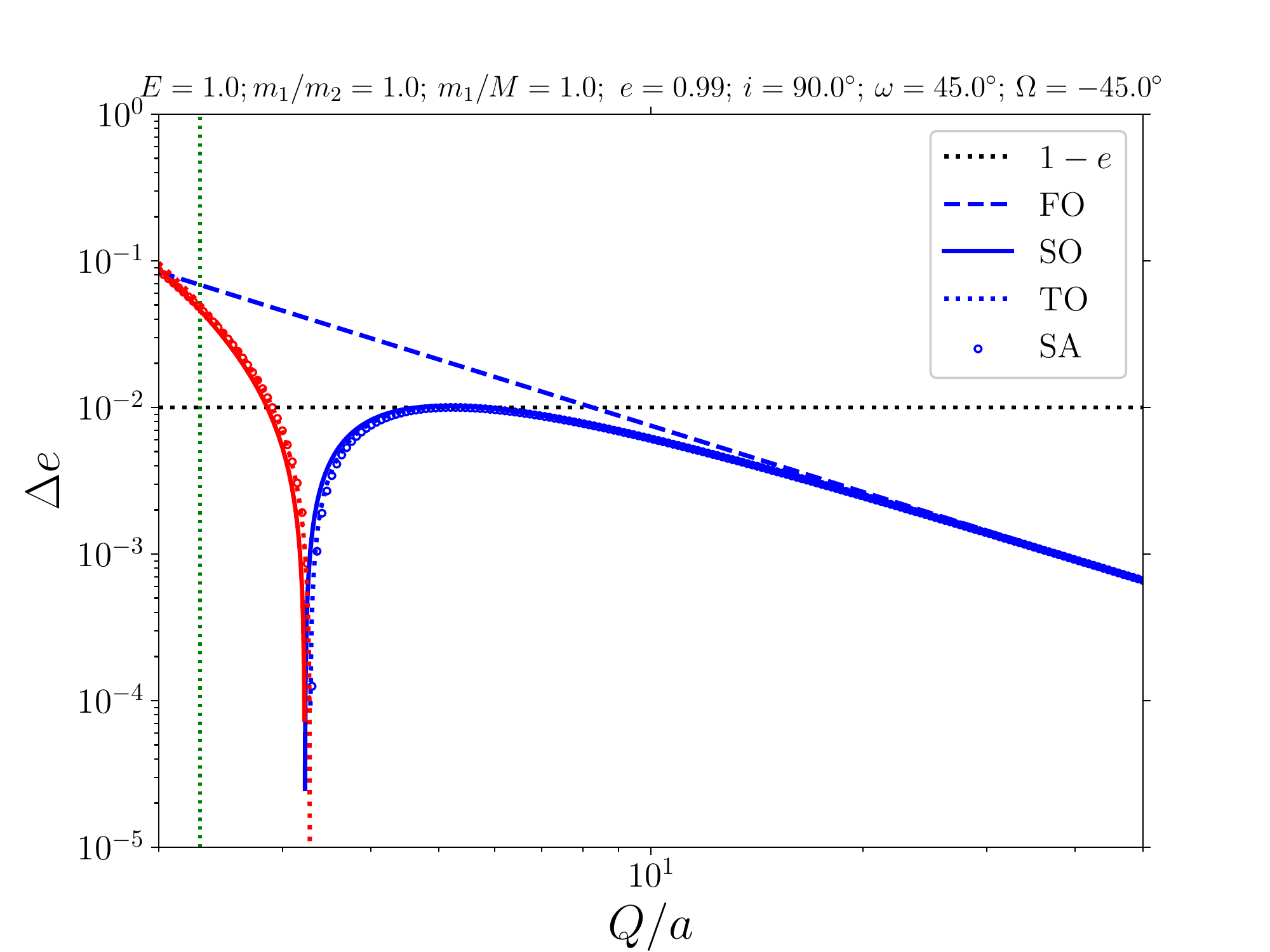}
\includegraphics[scale = 0.46, trim = 8mm 0mm 8mm 0mm]{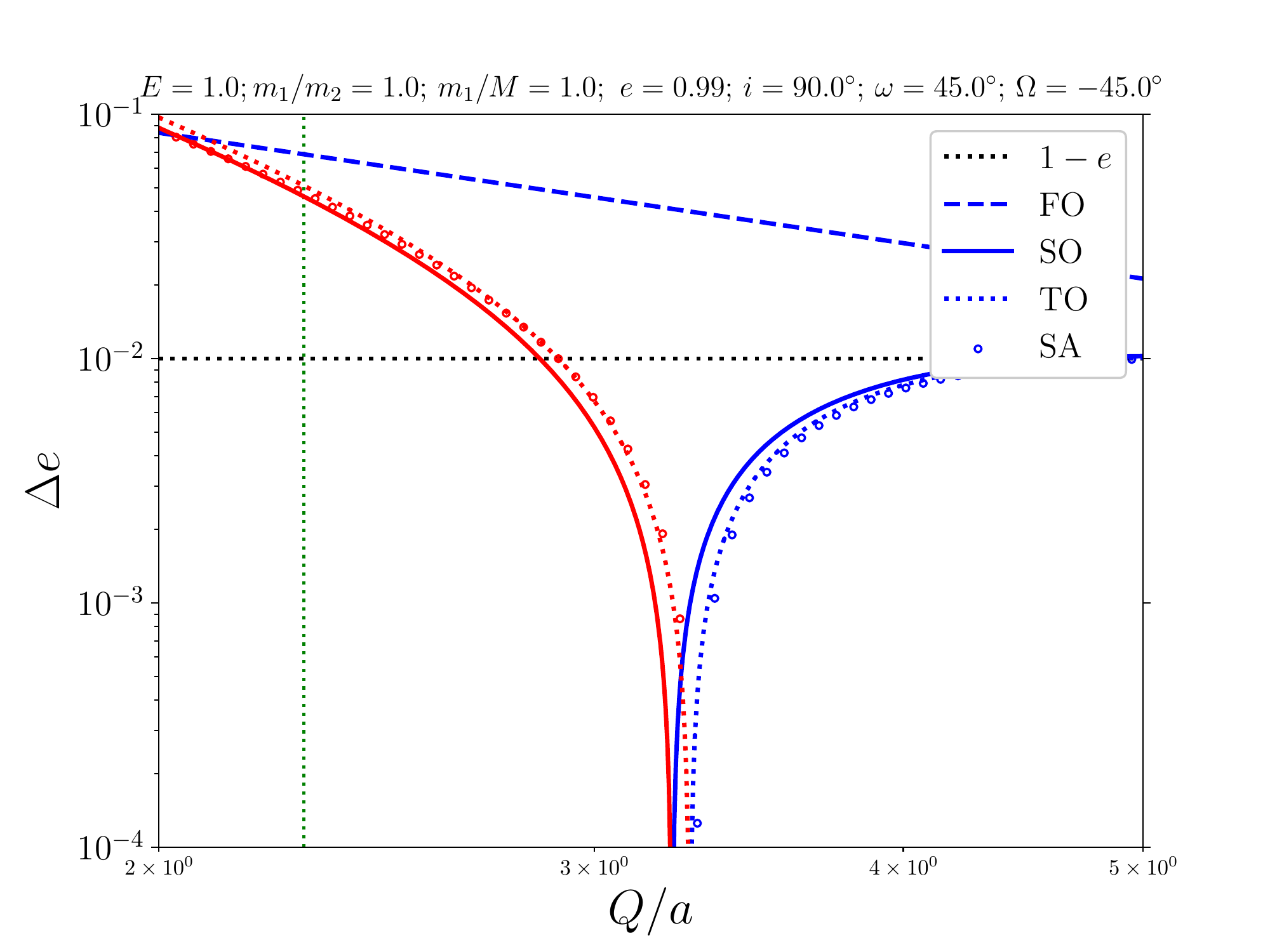}
\caption { An example of the scalar eccentricity change as a function of $Q/a$ according to numerically solving the SA equations of motion, equations~(\ref{eq:EOM_SA_gen}), and using analytic expressions: first order (FO; dashed lines), second order (SO; solid lines), and third order (TO; dotted lines) in $\epssa$ (refer to the legend). Blue (red) lines and symbols correspond to positive (negative) $\Delta e$. The FO, SO, and TO lines include octupole-order terms up to and including FO, SO, and TO, respectively, although in this case $m_1=m_2$ such that $\epsoct=0$. The right-hand panel is a zoomed-in version of the left-hand panel. The vertical dotted green line shows the value of $Q/a$ for which the `adiabatic ratio' of the angular speed of the perturber at periapsis to the binary mean motion (see equation 1 of \pI) is 0.5. To the left of this line, we expect the secular approximation to break down. }
\label{fig:test_TO}
\end{figure}

\begin{figure}
\center
\includegraphics[scale = 0.46, trim = 8mm 0mm 8mm 0mm]{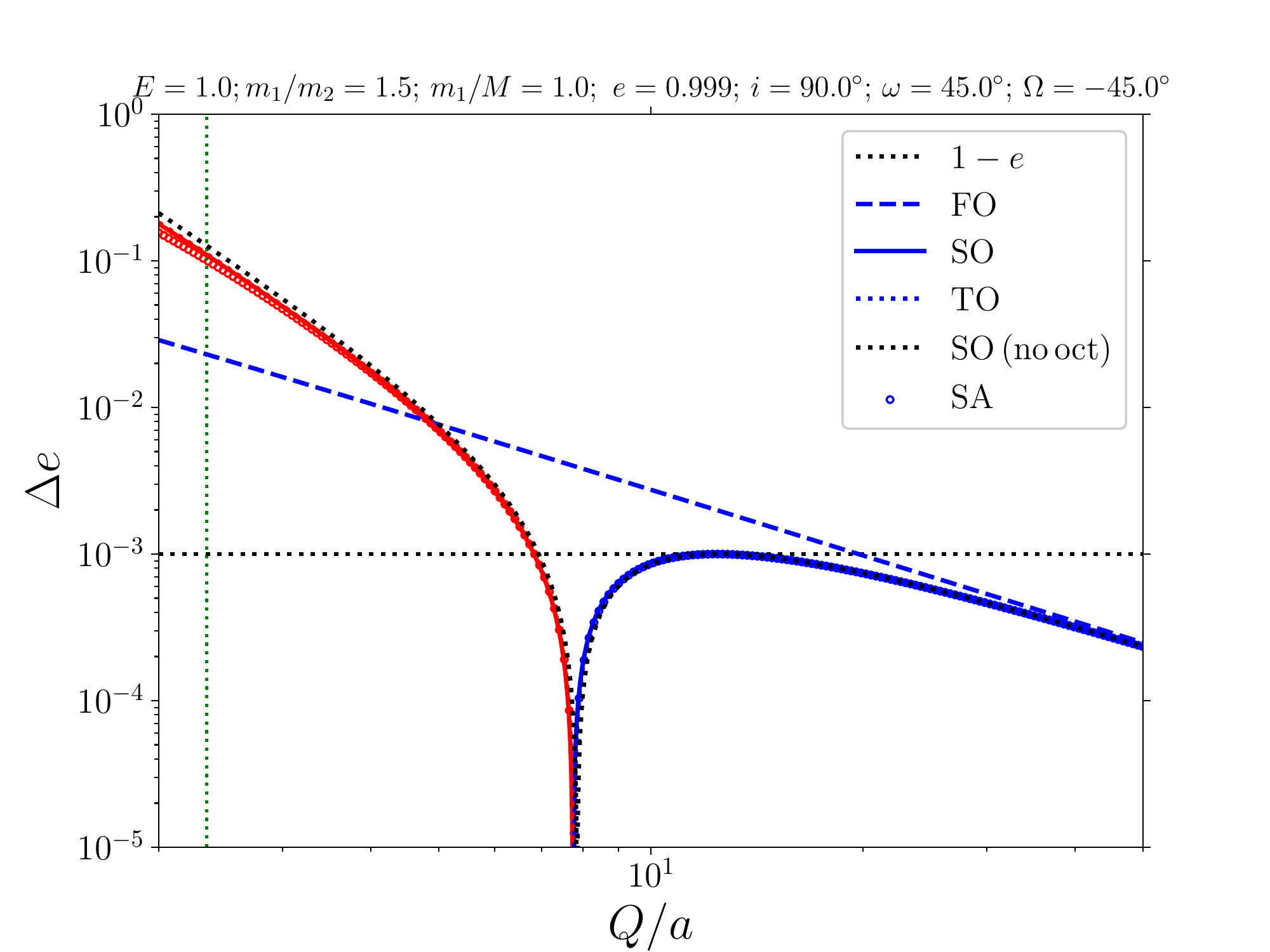}
\includegraphics[scale = 0.46, trim = 8mm 0mm 8mm 0mm]{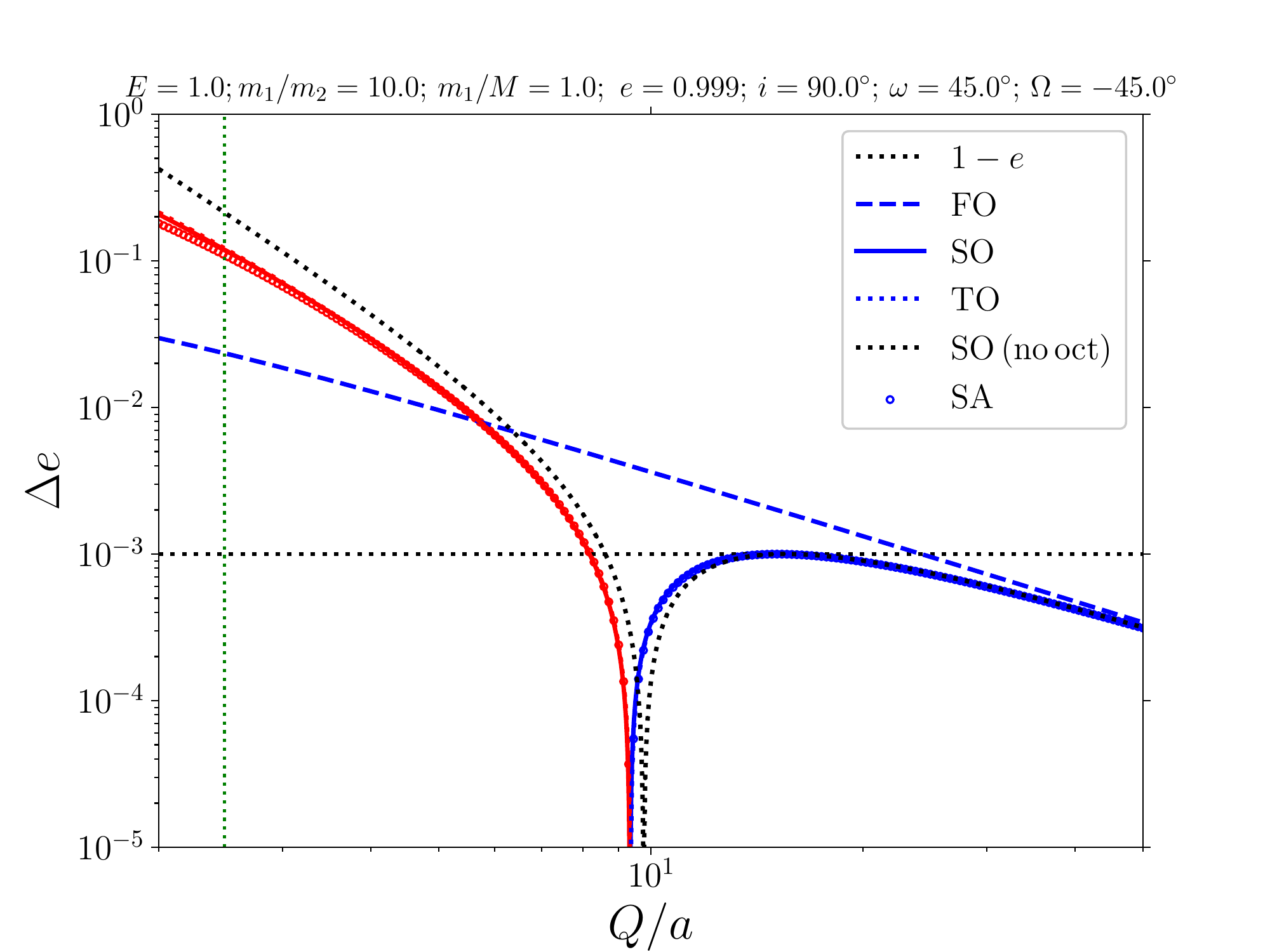}
\includegraphics[scale = 0.46, trim = 8mm 0mm 8mm 0mm]{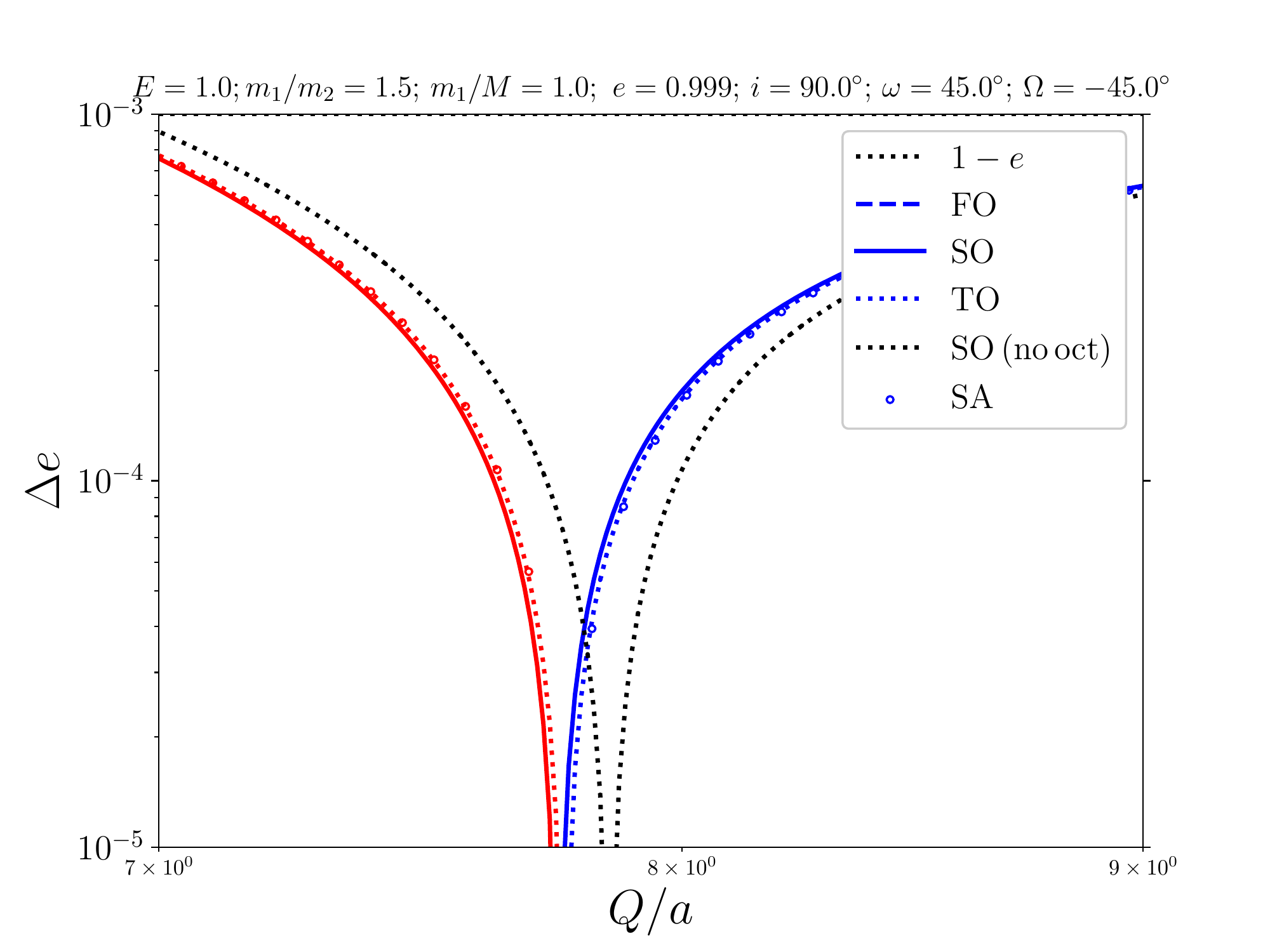}
\includegraphics[scale = 0.46, trim = 8mm 0mm 8mm 0mm]{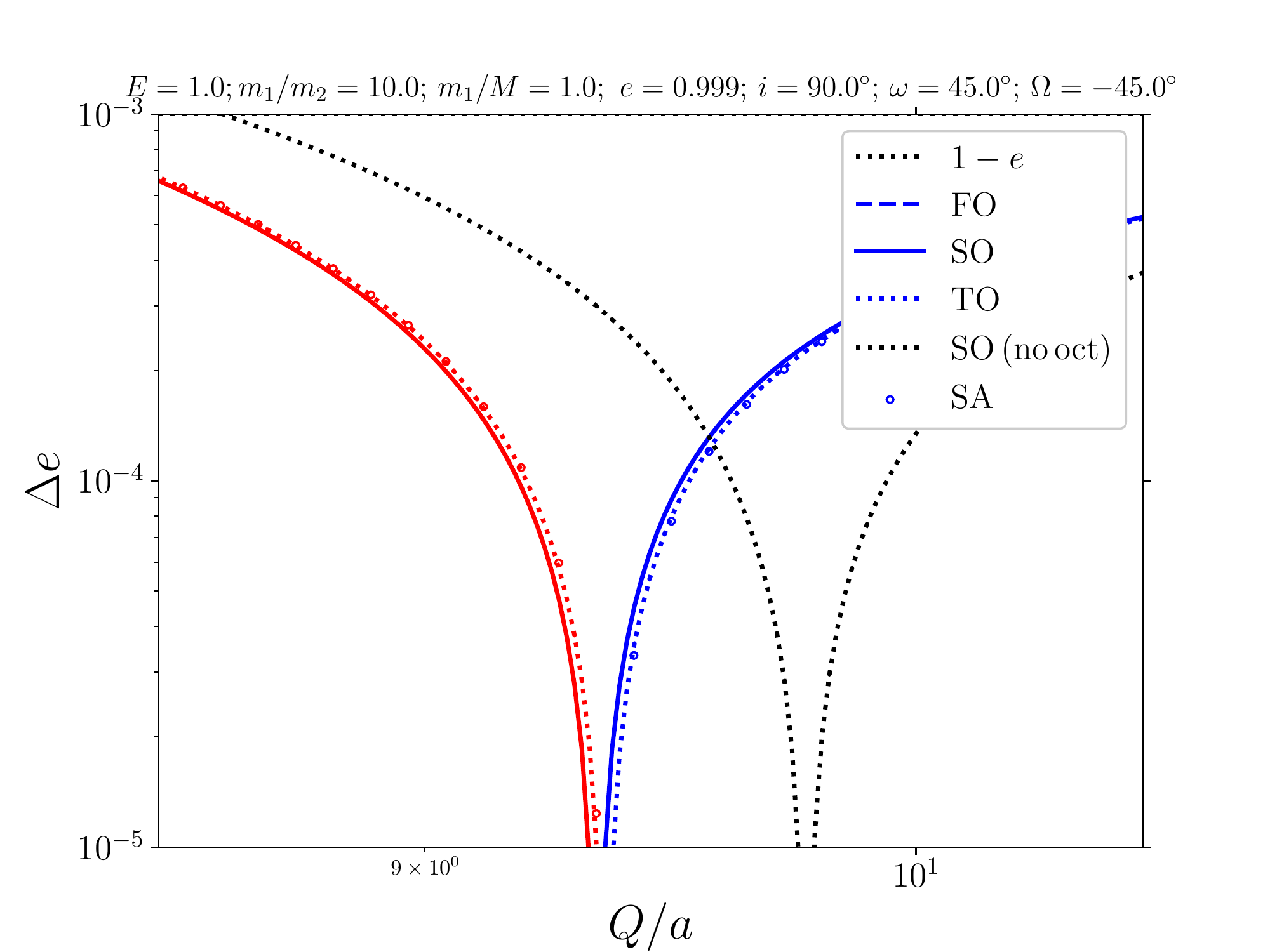}
\caption { Similar to \F~\ref{fig:test_TO}, here showing two examples with nonzero octupole-order terms -- the binary component masses are unequal; we set $m_1=\mper=1$ with $m_2=2/3$ and $m_2=1/10$ in the left and right-hand panels, respectively. The bottom panels are zoomed-in versions of the top panels. The (non-horizontal) black dotted lines show results from the analytic SO terms {\it without} the octupole-order terms derived in this paper. } 
\label{fig:test}
\end{figure}

In \F~\ref{fig:test_TO}, we show an example of the scalar eccentricity change $\Delta e$ as a function of $Q/a$ according to numerically solving the SA equations of motion, equations~(\ref{eq:EOM_SA_gen}), and using the analytic expressions: FO, SO, and TO in $\epssa$. The right-hand panel is a zoomed-in version of the left-hand panel. Here, we choose the same initial conditions as in the bottom-right-hand panel of Fig. 5 in \pI. In the latter panel in \pI, some deviation was found between the SA and SO results. Here, we also include the TO terms, and it can be seen that the addition of the TO terms solves the (minor) discrepancy. 

Next, we focus on the octupole-order terms. In \F~\ref{fig:test}, we adopt similar initial conditions as in Section 5.1 of \pI, i.e., with $m_1\neq m_2$, such that $\epsoct \neq 0$. In \pI, the octupole-order corrections were only included to FO in $\epssa$ (and, therefore, to FO in $\epsoct$). Consequently, there was significant deviation of the analytic prediction from the 3-body and SA integrations. Here, we include all octupole-order terms to the corresponding orders of $\epssa$, i.e., in \F~\ref{fig:test}, the FO, SO, and TO lines include octupole-order terms up to and including FO, SO, and TO in $\epsoct$, respectively. In the left (right)-hand panels, we set $m_1/m_2=1.5$ ($m_1/m_2=10$). The (non-horizontal) black dotted lines in \F~\ref{fig:test} show results from the analytic SO terms {\it without} the octupole-order terms derived in this paper, and illustrate that the octupole-order contribution can be important.  

As shown in \F~\ref{fig:test}, the additional octupole-order terms derived here give good agreement with the SA integrations. In particular, the value of $Q/a$ for which the sign change of $\Delta e$ occurs is accurately predicted. The bottom panels are zoomed-in versions of the top panels. In the bottom panels, it becomes clear that the TO terms give slightly better agreement with the SA integrations. The additional number of terms introduced by the TO terms is significant, however, as shown in Table~\ref{table:N_terms}.

\subsubsection{Further exploration of parameter space}
\label{sect:TO:test_ex:ex}

\begin{figure}
\center
\includegraphics[scale = 0.46, trim = 8mm 0mm 8mm 0mm]{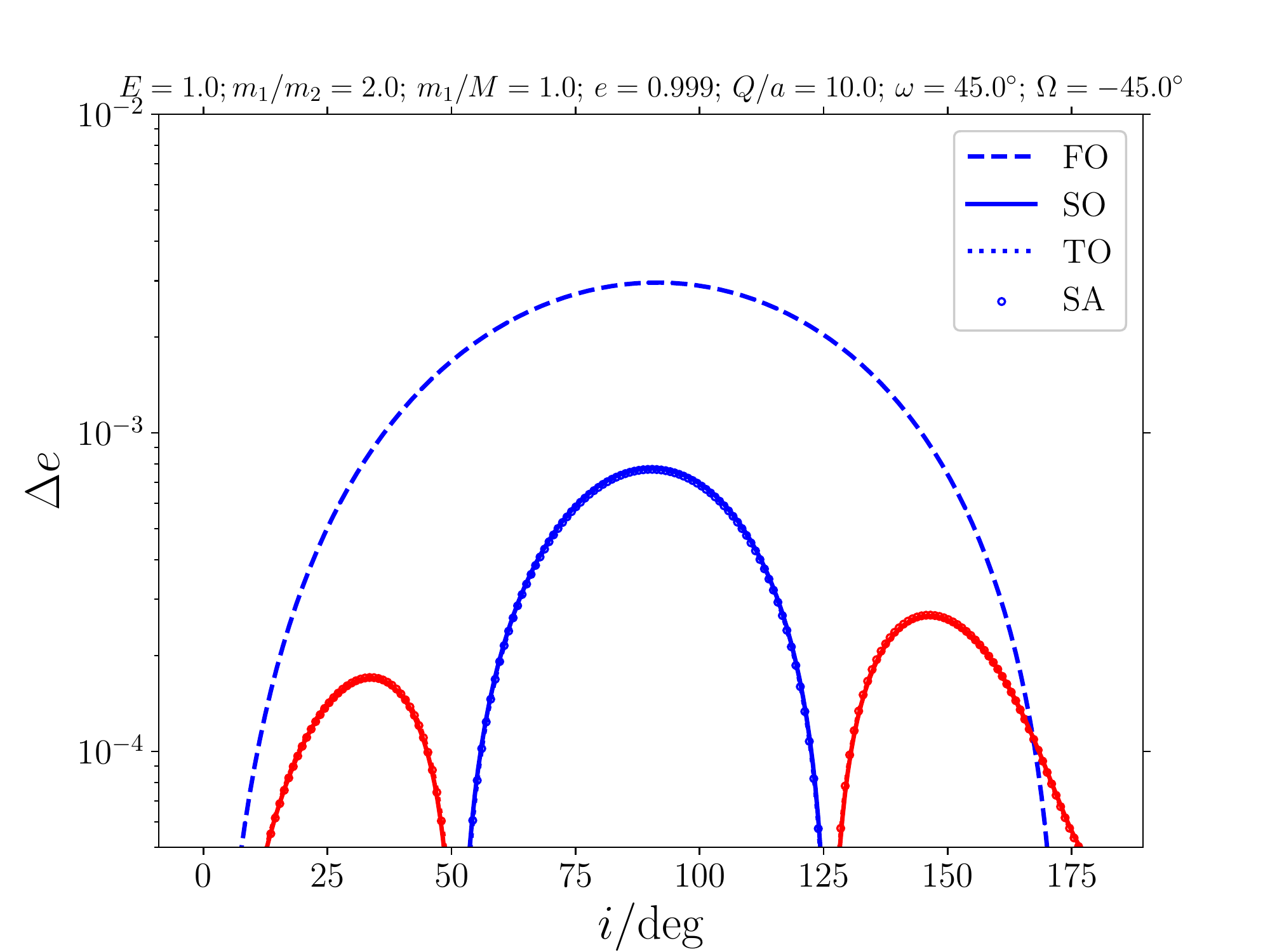}
\includegraphics[scale = 0.46, trim = 8mm 0mm 8mm 0mm]{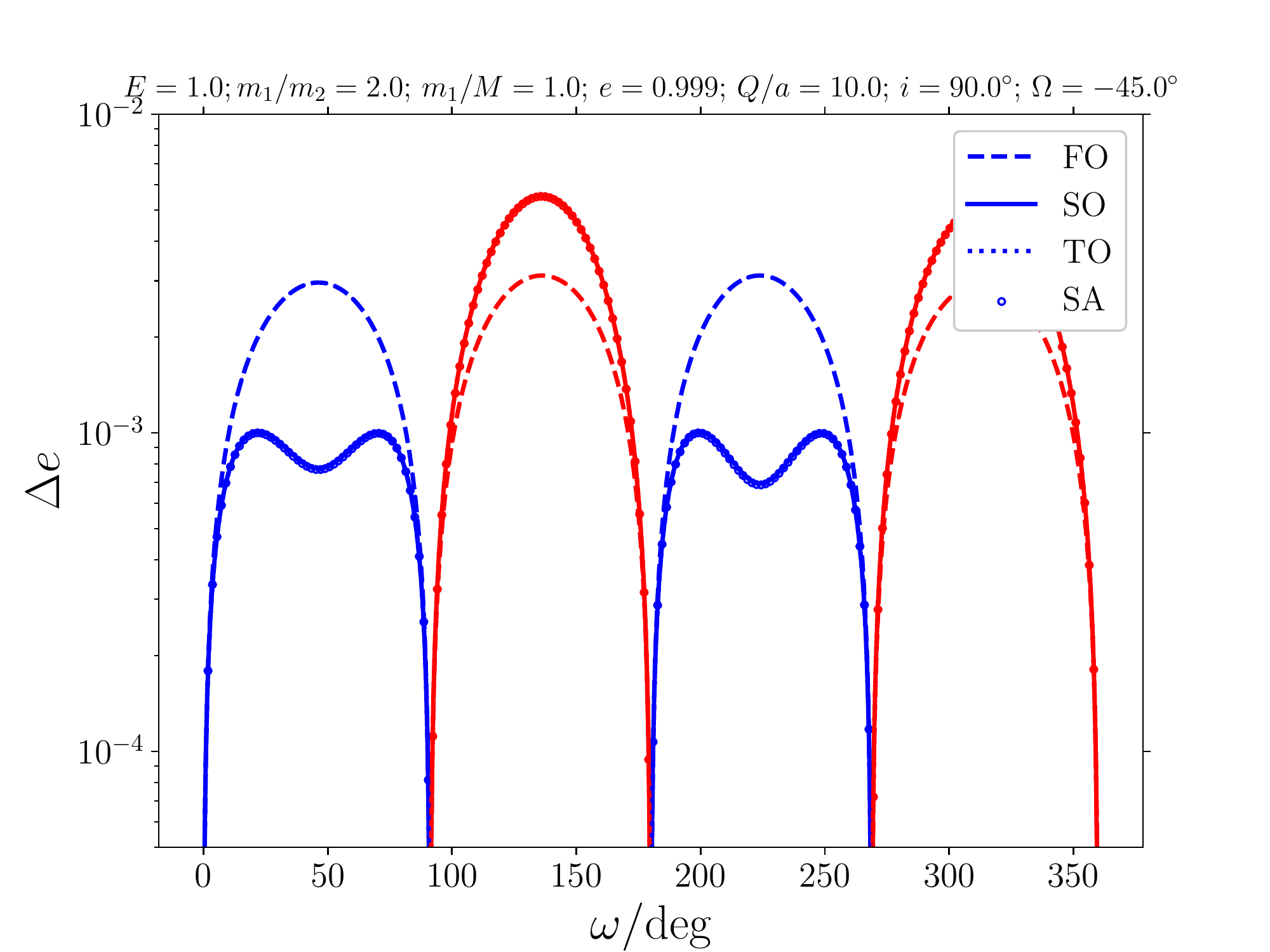}
\caption { Scalar eccentricity change as a function of $i$ (left-hand panel), and of $\omega$ (right-hand panel). Blue (red) colors correspond to positive (negative) $\Delta e$. The dependence on $\Omega$ is very weak, and is not shown. The fixed parameters assumed in each panel are indicated in the top of each panel. }
\label{fig:ex1}
\end{figure}

\begin{figure}
\center
\includegraphics[scale = 0.46, trim = 8mm 0mm 8mm 0mm]{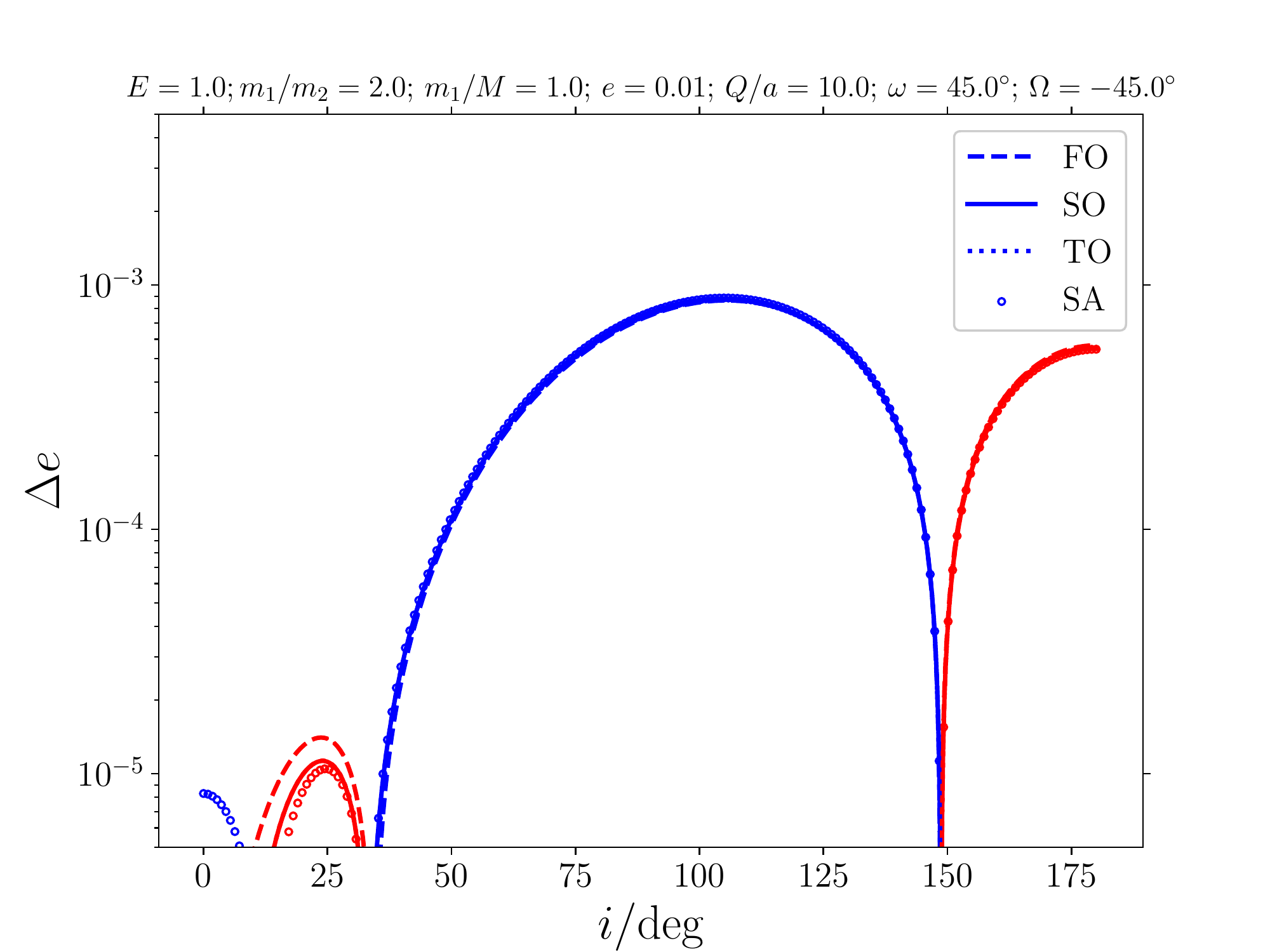}
\includegraphics[scale = 0.46, trim = 8mm 0mm 8mm 0mm]{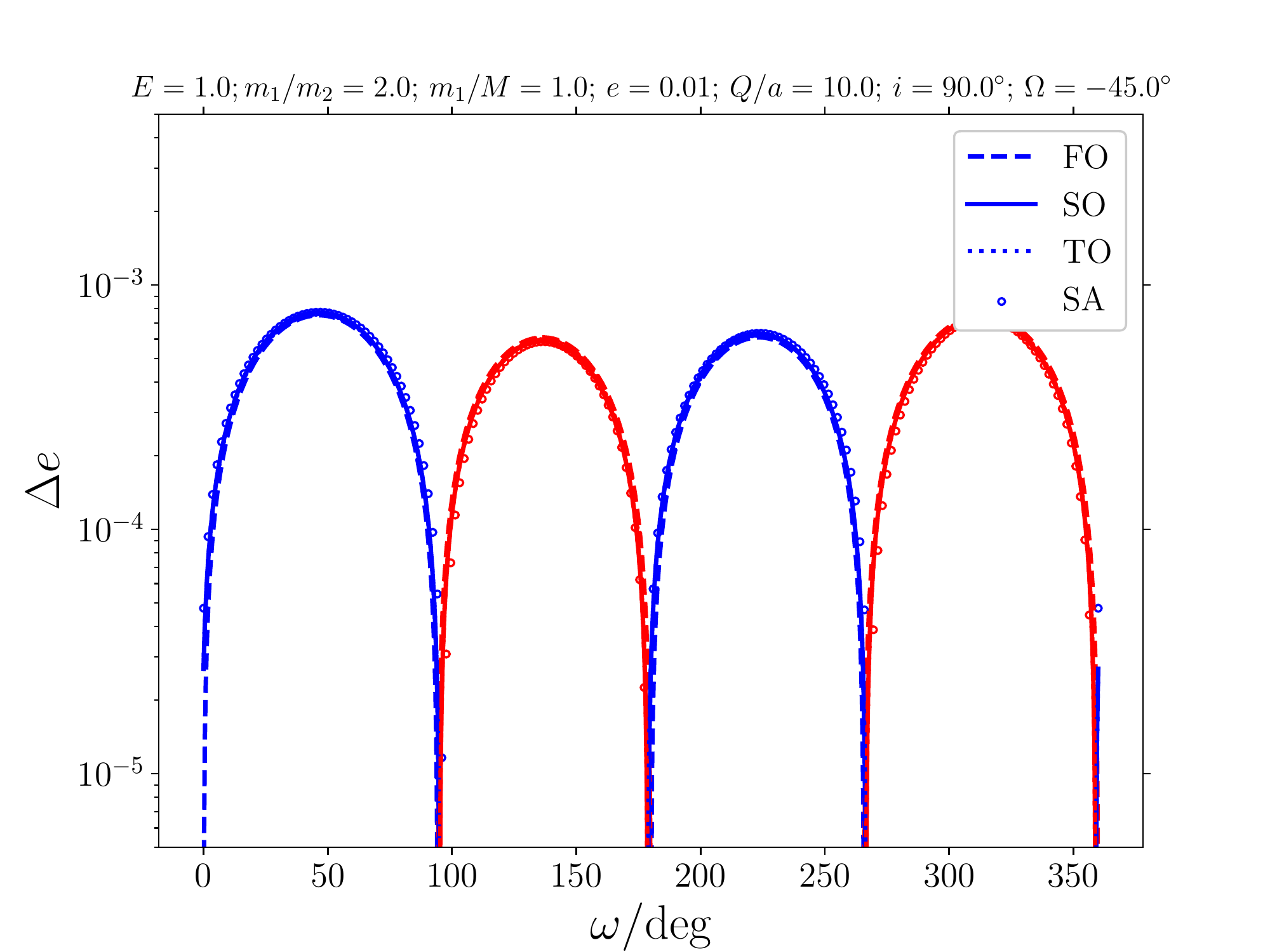}
\caption{ Similar to \F~\ref{fig:ex1}, here assuming a low initial eccentricity of $e=0.01$.}
\label{fig:ex2}
\end{figure}

\begin{figure}
\center
\includegraphics[scale = 0.46, trim = 8mm 0mm 8mm 0mm]{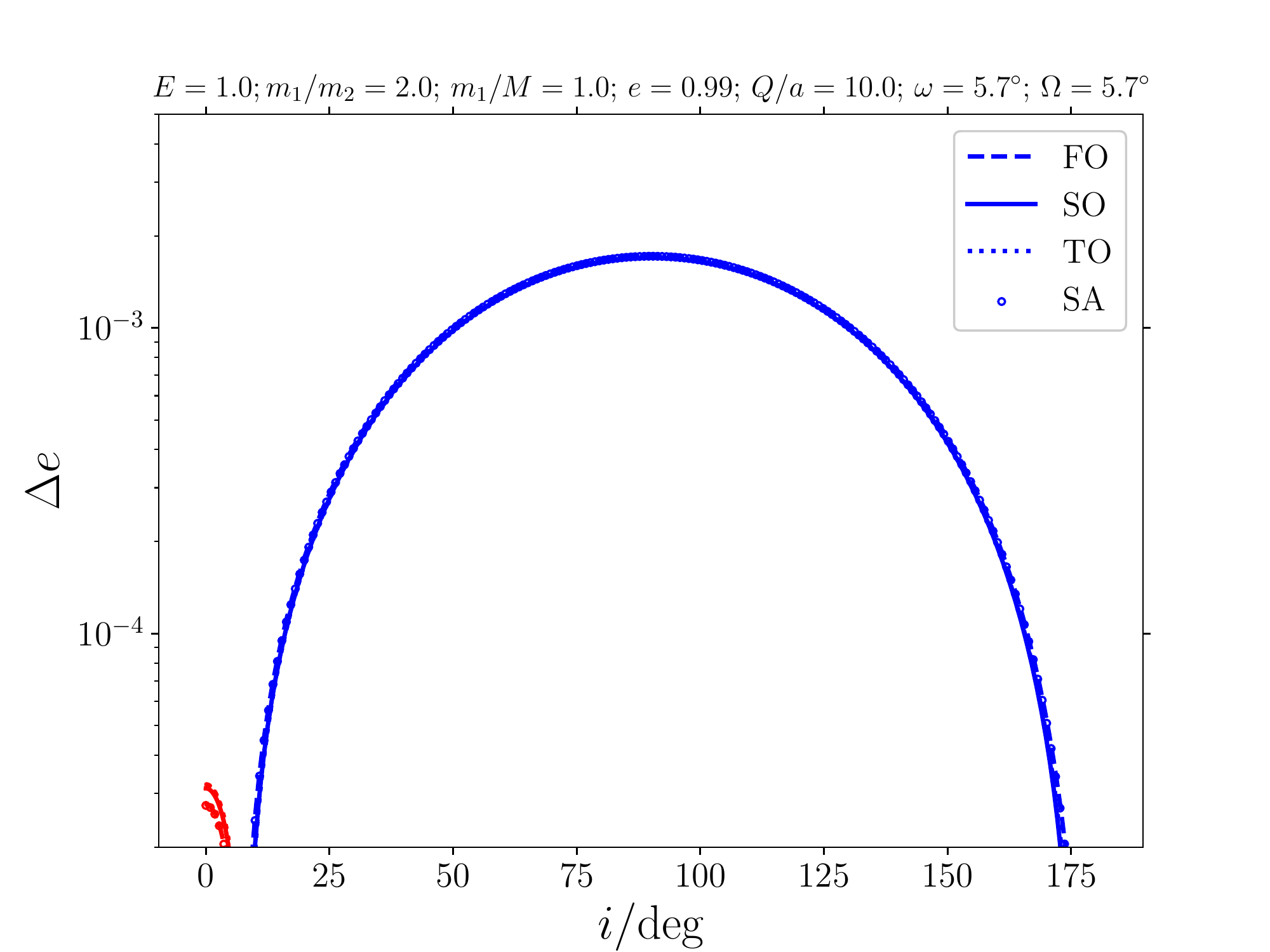}
\includegraphics[scale = 0.46, trim = 8mm 0mm 8mm 0mm]{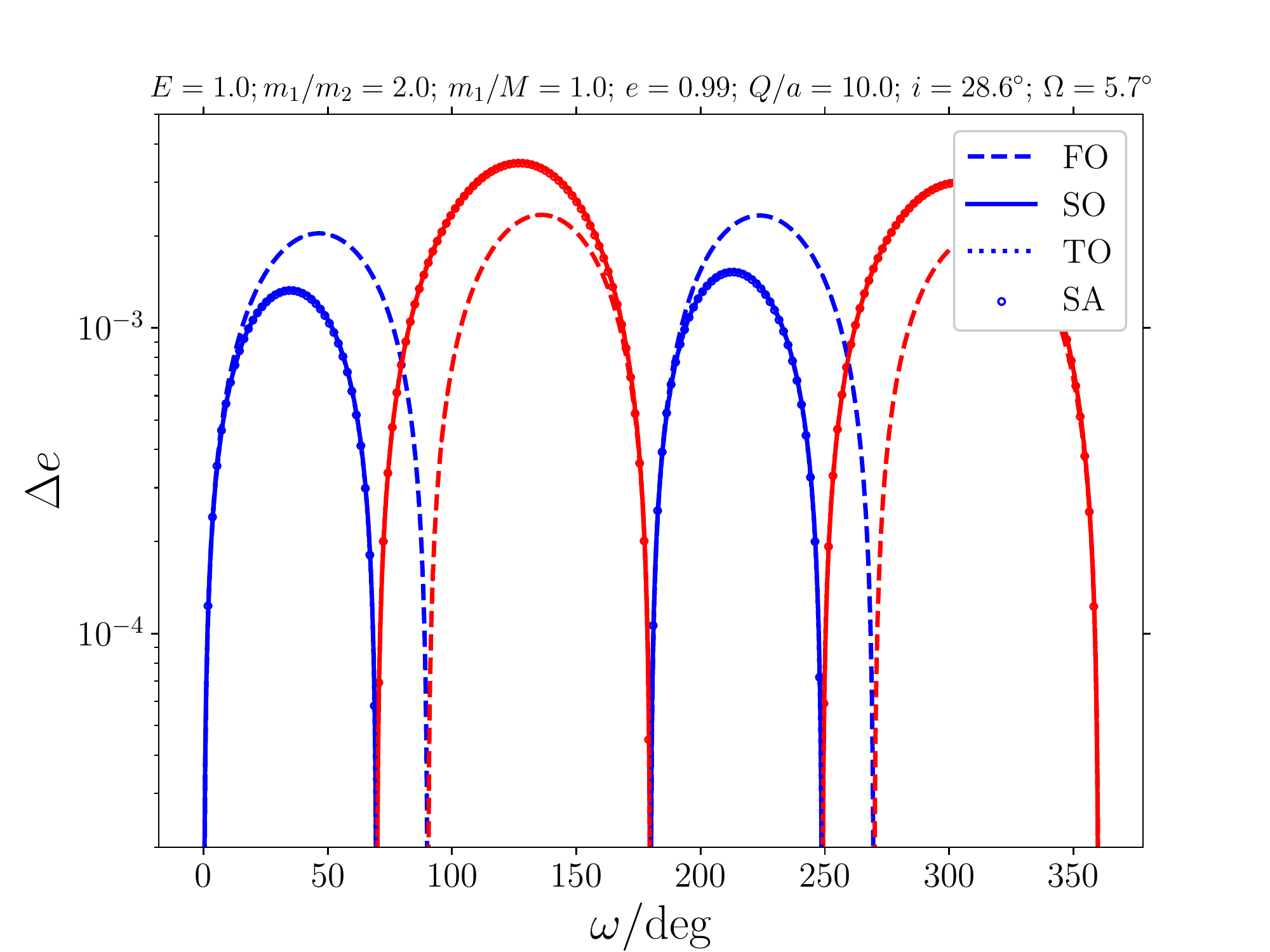}
\includegraphics[scale = 0.46, trim = 8mm 0mm 8mm 0mm]{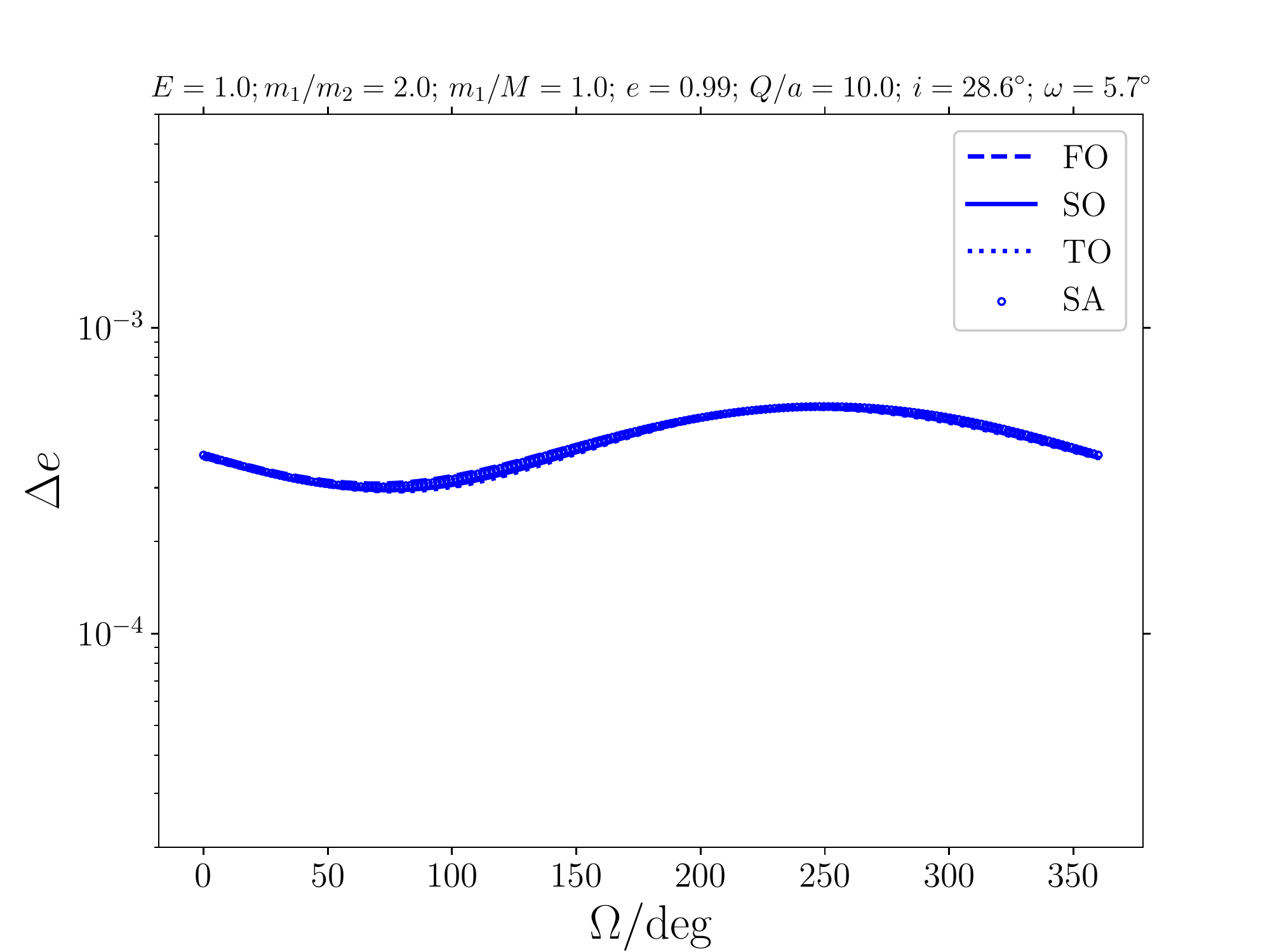}
\caption{ Similar to \F~\ref{fig:ex1}, here assuming different fixed values of $\omega$ and $\Omega$. }
\label{fig:ex3}
\end{figure}

Here, we explore in more detail the dependence of $\Delta e$ on the assumed parameters. In \F~\ref{fig:ex1}, we plot $\Delta e$ as a function of the inclination $i$ (left-hand panel) and the argument of periapsis $\omega$ (right-hand panel), for fixed other parameters (the dependence on the longitude of the ascending node $\Omega$ for the fixed other parameters is very weak, and not shown). For the chosen fixed parameters, $\Delta e$ is positive for $i$ around $90^\circ$, whereas $\Delta e<0$ if $i$ is close to 0 or $180^\circ$. This sign change is not captured by the FO expression, which predicts positive $\Delta e$ throughout (specifically, $\Delta e \propto \sin^2 i \geq 0$). Considered as a function of $\omega$, $\Delta e$ changes sign multiple times. The FO expression predicts a simple dependence, $\Delta e \propto \sin 2 \omega$, which approximately captures the dependence on $\omega$. However, the numerical SA integrations show a more elaborate behaviour which is well captured by the SO and TO expressions.

In \F~\ref{fig:ex2}, the initial eccentricity is assumed to be $e=0.01$, in contrast to $e=0.999$ in \F~\ref{fig:ex1}. The inclination dependence (left-hand panel) is now asymmetric, and this behaviour is captured by the FO, SO, and TO expressions. The detailed features as a function of $\omega$ in \F~\ref{fig:ex1} are not apparent in the lower-eccentricity example in \F~\ref{fig:ex2}. 

In \F~\ref{fig:ex3}, we set $e=0.99$, and choose different values for the fixed $\omega$ and $\Omega$. The scalar eccentricity change is positive for most inclinations except for small inclinations, which is captured by all analytic methods. Considered as a function of $\omega$, $\Delta e$ shows cyclical behaviour, which is not entirely captured by the FO expression, but accurately with the higher-order expressions. For these parameters, the dependence of $\Delta e$ on $\Omega$ is no longer very weak, and is shown in the third panel of \F~\ref{fig:ex3}. The dependence is sinusoidal, and is well captured by all analytic expressions.

\section{Post-Newtonian terms}
\label{sect:PN}
Here, we consider in more detail compared to \pI~the effects of post-Newtonian (PN) terms. We restrict to the lowest-order non-dissipative PN terms, i.e., those associated with $c^2$, where $c$ is the speed of light. As is well known, averaged over the mean motion of the binary, the 1PN terms give rise to precession of $\ve{e}$ around $\ve{\j}$, described by \citep{1972gcpa.book.....W}
\begin{align}
\label{eq:1PN_t}
\frac{\mathrm{d} \ve{e}}{\mathrm{d} t} = 3 n_\bin \frac{\rg}{a} \frac{1}{1-e^2} e \, \unit{q},
\end{align}
where $\rg \equiv Gm/c^2$ is the binary's gravitational radius, $n_\bin\equiv \sqrt{Gm/a^3}$ is the binary's mean motion, and $\unit{q} \equiv \unit{\j} \times \unit{e}$ is a unit vector perpendicular to both $\unit{e}$ and $\unit{\j}$. Formulated in terms of the perturber's true anomaly $\theta$ (and, again, averaged over the binary's mean motion), 1PN precession give rise to an additional term to the equations of motion, equations~(\ref{eq:EOM_SA_gen}), which is given by
\begin{align}
\label{eq:1PN}
\left. \frac{\mathrm{d} \ve{e}}{\mathrm{d} \theta}\right |_{\mathrm{1PN}} = \epsopn \frac{e \, \unit{q}}{1-e^2} \frac{1}{(1+\eper \cos \theta)^2},
\end{align}
where
\begin{align}
\epsopn \equiv 3 \frac{\rg}{a} \left [ \frac{m}{m+\mper} \left ( \frac{Q}{a} \right )^3 (1+\eper)^3 \right ]^{1/2}.
\end{align}
From equation~(\ref{eq:1PN}), it is immediately clear that the 1PN term $\mathrm{d} \ve{e}/\mathrm{d}\theta $ diverges when integrating it from $\theta=-\Leper$ to $\theta=\Leper$. Physically, this means that the amount of 1PN-induced precession is infinite over a time span corresponding to $t\rightarrow -\infty$ to $t\rightarrow \infty$. Unfortunately, this implies that the procedure used in \pI~to compute Fourier coefficients based on the equations of motion directly cannot straightforwardly be extended to include the 1PN terms directly in the equations of motion.

Another important implication is that the infinite precession of the binary corresponding to $-\Leper<\theta<\Leper$ implies that the orientation of the binary when the perturber passes it is ill-defined. Consequently, the 1PN terms give rise to an uncertainty in the problem -- the initial argument of periapsis of the binary can no longer be considered to be a true initial parameter. Therefore, we will consider the {\it statistical} properties of the eccentricity changes over an ensemble of the initial arguments of periapsis, $\omega_{\mathrm{i}}$, where we take the initial distribution of $\omega_{\mathrm{i}}$ to be uniform between 0 and $2\pi$.

The Newtonian predictions for the eccentricity change in this case can be straightforwardly computed from the analytic expressions derived in \pI~and in \S~\ref{sect:TO}. Specifically, by expressing the scalar eccentricity change in terms of orbital elements and averaging over $\omega$, we find that the mean and root-mean-squared (rms) scalar eccentricity change to TO in $\epssa$ are given by (setting $\epsoct=0$ and $\eper=1$ for simplicity)
\begin{subequations}
\label{eq:stat_N}
\begin{align}
\nonumber \langle \Delta e \rangle_\omega &\equiv \int_{0}^{2\pi} \mathrm{d} \omega \, \Delta e = \frac{9}{512} \pi ^2  \epssa^2 e \left[4 \left(81 e^2-56\right) \cos (2 i)+\left(39 e^2+36\right) \cos (4 i)-299 e^2+124\right] \\
\nonumber &\quad -\frac{9}{4096} \pi  \epssa^3 e  \sqrt{1-e^2} \cos (i) \Biggl [176 e^2 \sin (2 i-4 \Omega )-44 e^2 \sin (4 i-4 \Omega )+854 e^2 \sin (2 i-2 \Omega )-1001 e^2 \sin (4 i-2 \Omega )  \\
\nonumber &\qquad  -854 e^2 \sin (2 (i+\Omega ))+44 e^2 \sin (4 (i+\Omega ))+1001 e^2 \sin (4 i+2 \Omega )-176 e^2 \sin (2 i+4 \Omega )-880 \pi  e^2 \cos (2 i-2 \Omega )  \\
\nonumber &\qquad  +660 \pi  e^2 \cos (4 i-2 \Omega )-880 \pi  e^2 \cos (2 (i+\Omega ))+660 \pi  e^2 \cos (4 i+2 \Omega )+40 \pi  \left(49 e^2-4\right) \cos (2 i)-120 \pi  \left(11 e^2+4\right) \cos (4 i)  \\
\nonumber &\qquad  -294 e^2 \sin (2 \Omega )+264 e^2 \sin (4 \Omega )+440 \pi  e^2 \cos (2 \Omega )+3360 \pi  e^2+64 \sin (2 i-4 \Omega )-16 \sin (4 i-4 \Omega )+616 \sin (2 i-2 \Omega )  \\
\nonumber &\qquad  -364 \sin (4 i-2 \Omega )-616 \sin (2 (i+\Omega ))+16 \sin (4 (i+\Omega ))+364 \sin (4 i+2 \Omega )-64 \sin (2 i+4 \Omega )-320 \pi  \cos (2 i-2 \Omega ) \\
&\qquad +240 \pi  \cos (4 i-2 \Omega )-320 \pi  \cos (2 (i+\Omega )) + 240 \pi  \cos (4 i+2 \Omega )+504 \sin (2 \Omega )+96 \sin (4 \Omega )+160 \pi  \cos (2 \Omega )+640 \pi \Biggl ]; \\
\nonumber \left \langle (\Delta e)^2 \right \rangle^{1/2}_\omega &\equiv \left [ \int_{0}^{2\pi} \mathrm{d} \omega \, (\Delta e)^2 \right ]^{1/2} = \frac{15}{4 \sqrt{2}} \pi  \epssa e \sqrt{1-e^2} \sin ^2(i) \\
\nonumber &\quad + \frac{75}{128 \sqrt{2}} \pi  \epssa^2 e \left(1-e^2\right) \cos (i) (3 \cos (2 i-2 \Omega )+3 \cos (2 (i+\Omega ))-6 \cos (2 i)+2 \cos (2 \Omega )+6) \\
\nonumber &\quad -\frac{3}{1310720 \sqrt{2}}\pi \epssa^3 \frac{e}{\sqrt{1-e^2}}  \csc ^2(i) \Biggl [ -348456 \pi ^2 \cos (6 i) e^4-51100 \cos (6 i) e^4-69867 \pi ^2 \cos (8 i) e^4-63800 \cos (2 i-4 \Omega ) e^4 \\
\nonumber &\qquad +232560 \cos (4 i-4 \Omega ) e^4-25160 \cos (6 i-4 \Omega ) e^4-607460 \cos (2 i-2 \Omega ) e^4+19880 \cos (4 i-2 \Omega ) e^4-232540 \cos (6 i-2 \Omega ) e^4 \\
\nonumber &\qquad +1640240 \cos (2 \Omega ) e^4+352800 \cos (4 \Omega ) e^4-607460 \cos (2 (i+\Omega )) e^4+232560 \cos (4 (i+\Omega )) e^4+19880 \cos (4 i+2 \Omega ) e^4\\
\nonumber &\qquad -232540 \cos (6 i+2 \Omega ) e^4-63800 \cos (2 i+4 \Omega ) e^4-25160 \cos (6 i+4 \Omega ) e^4-82800 \pi  \sin (2 i-2 \Omega ) e^4+309600 \pi  \sin (4 i-2 \Omega ) e^4\\
\nonumber &\qquad -13200 \pi  \sin (6 i-2 \Omega ) e^4+427200 \pi  \sin (2 \Omega ) e^4+82800 \pi  \sin (2 (i+\Omega )) e^4-309600 \pi  \sin (4 i+2 \Omega ) e^4+13200 \pi  \sin (6 i+2 \Omega ) e^4\\
\nonumber &\qquad -8084769 \pi ^2 e^4+241400 e^4-228288 \pi ^2 \cos (6 i) e^2+7700 \cos (6 i) e^2-92016 \pi ^2 \cos (8 i) e^2+53800 \cos (2 i-4 \Omega ) e^2 \\
\nonumber &\qquad -375120 \cos (4 i-4 \Omega ) e^2+18520 \cos (6 i-4 \Omega ) e^2+503020 \cos (2 i-2 \Omega ) e^2+73640 \cos (4 i-2 \Omega ) e^2+202580 \cos (6 i-2 \Omega ) e^2\\
\nonumber &\qquad-1558480 \cos (2 \Omega ) e^2-674400 \cos (4 \Omega ) e^2+503020 \cos (2 (i+\Omega )) e^2-375120 \cos (4 (i+\Omega )) e^2+73640 \cos (4 i+2 \Omega ) e^2\\
\nonumber &\qquad +202580 \cos (6 i+2 \Omega ) e^2+53800 \cos (2 i+4 \Omega ) e^2+18520 \cos (6 i+4 \Omega ) e^2+27600 \pi  \sin (2 i-2 \Omega ) e^2-247200 \pi  \sin (4 i-2 \Omega ) e^2\\
\nonumber &\qquad+68400 \pi  \sin (6 i-2 \Omega ) e^2-302400 \pi  \sin (2 \Omega ) e^2-27600 \pi  \sin (2 (i+\Omega )) e^2+247200 \pi  \sin (4 i+2 \Omega ) e^2-68400 \pi  \sin (6 i+2 \Omega ) e^2\\
\nonumber &\qquad +7626288 \pi ^2 e^2+210200 e^2+12 \left(\left(-69475+871678 \pi ^2\right) e^4+\left(48825-948656 \pi ^2\right) e^2+185728 \pi ^2+20650\right) \cos (2 i)\\
\nonumber &\qquad -100 \left\{\left(-34+21045 \pi ^2\right) e^4-2 \left(2381+21864 \pi ^2\right) e^2+13008 \pi ^2+4796\right\} \cos (4 i)+351744 \pi ^2 \cos (6 i)+43400 \cos (6 i)\\
\nonumber &\qquad -34992 \pi ^2 \cos (8 i)+10000 \cos (2 i-4 \Omega )+142560 \cos (4 i-4 \Omega )+6640 \cos (6 i-4 \Omega )+104440 \cos (2 i-2 \Omega )\\
\nonumber &\qquad -93520 \cos (4 i-2 \Omega )+29960 \cos (6 i-2 \Omega )-81760 \cos (2 \Omega )+321600 \cos (4 \Omega )+104440 \cos (2 (i+\Omega ))+142560 \cos (4 (i+\Omega ))\\
\nonumber &\qquad -93520 \cos (4 i+2 \Omega )+29960 \cos (6 i+2 \Omega )+10000 \cos (2 i+4 \Omega )+6640 \cos (6 i+4 \Omega )+55200 \pi  \sin (2 i-2 \Omega )\\
\nonumber &\qquad -62400 \pi  \sin (4 i-2 \Omega )-55200 \pi  \sin (6 i-2 \Omega )-124800 \pi  \sin (2 \Omega )-55200 \pi  \sin (2 (i+\Omega ))+62400 \pi  \sin (4 i+2 \Omega ) \\
&\qquad +55200 \pi  \sin (6 i+2 \Omega )-1392144 \pi ^2-451600 \Biggl ].
\end{align}
\end{subequations}
Note that, when only the FO terms are included, $\langle \Delta e \rangle_{\omega} = 0$. 

\begin{figure}
\center
\includegraphics[scale = 0.46, trim = 8mm 0mm 8mm 0mm]{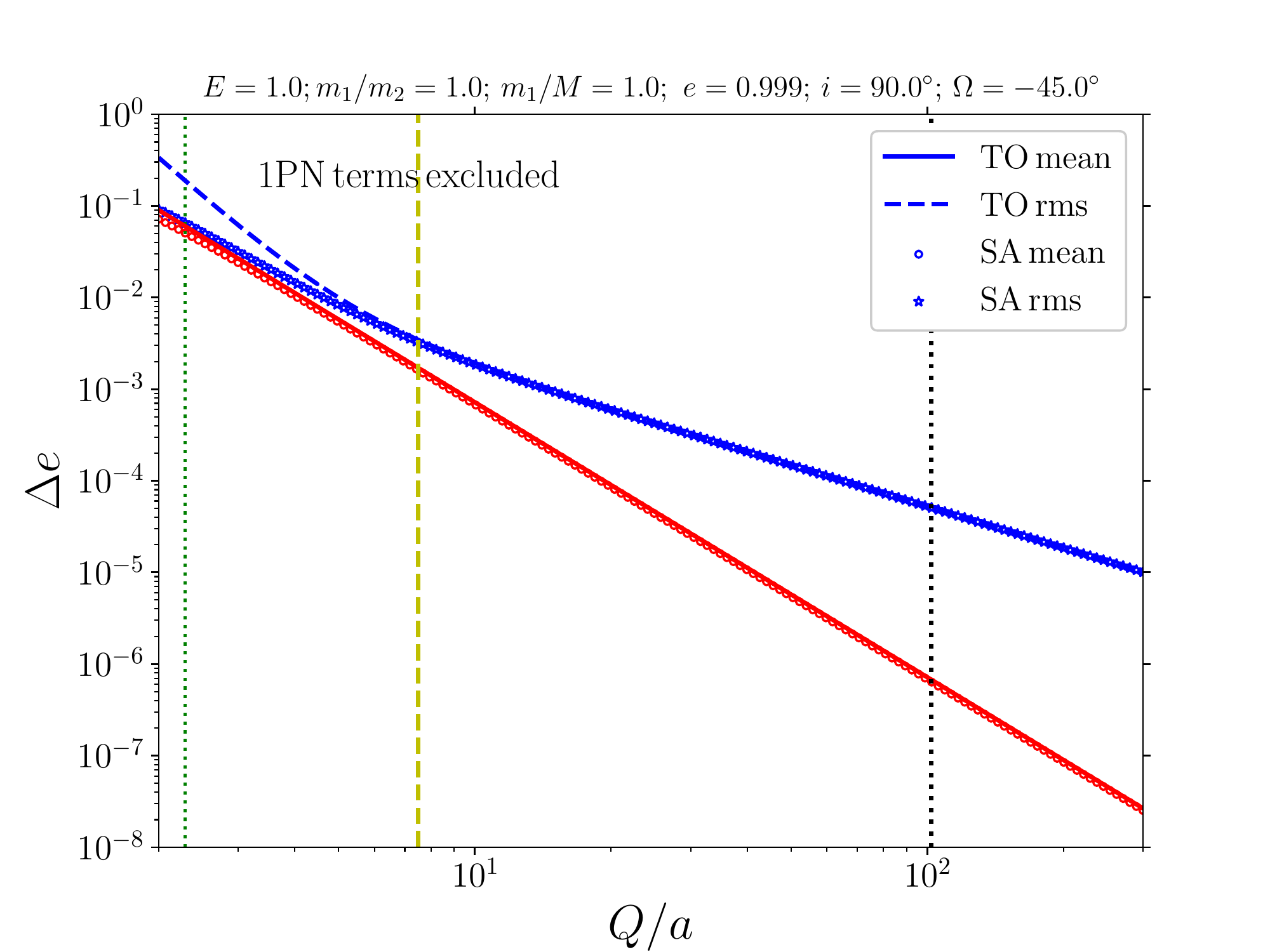}
\includegraphics[scale = 0.46, trim = 8mm 0mm 8mm 0mm]{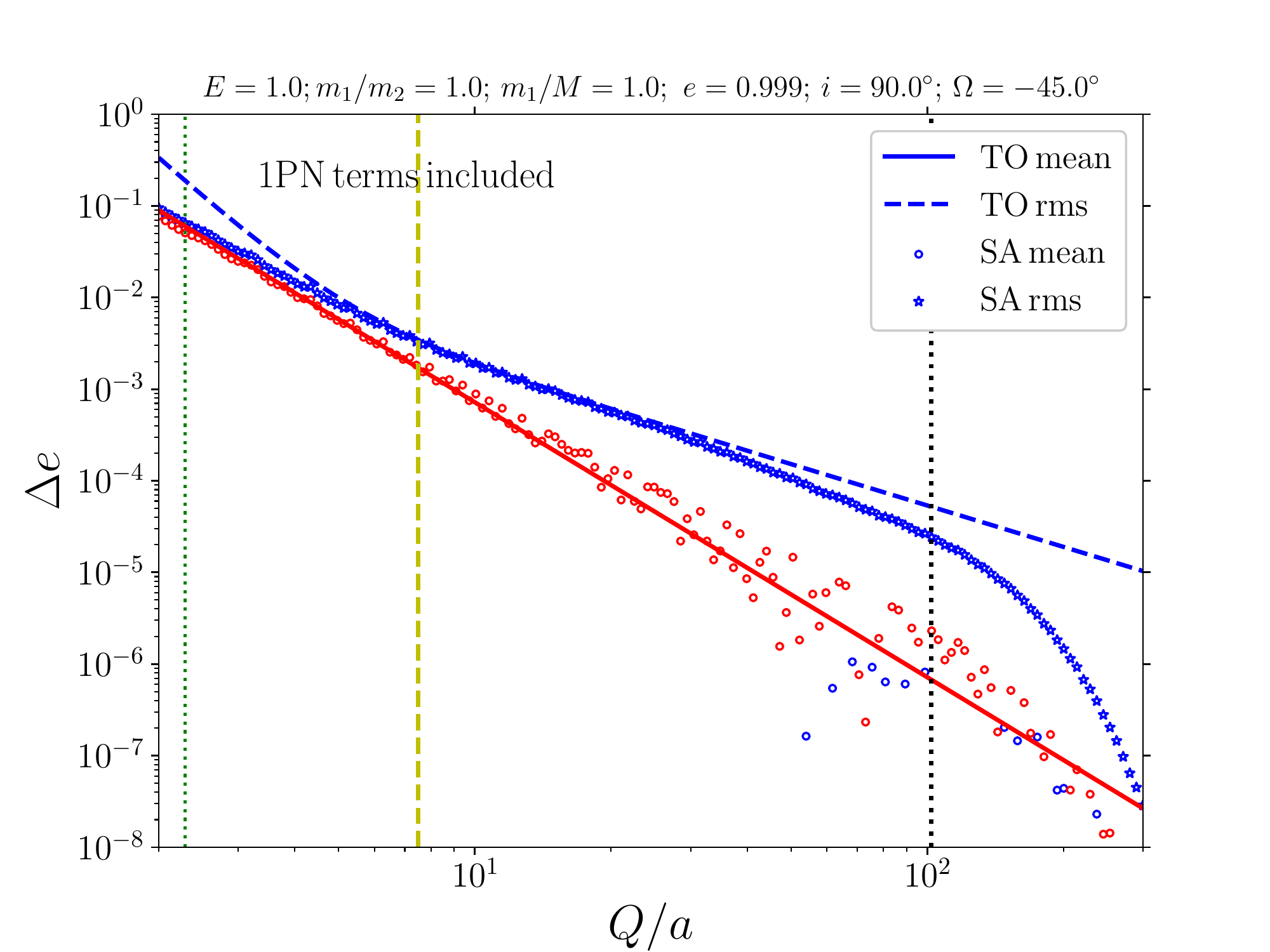}
\caption{  Mean and rms scalar eccentricity changes as a function of $Q/a$ (keeping $a$ fixed while varying $Q$) for a system with $m_1=m_2=\mper=40\,\msun$, $a=1\,\au$, $e=0.999$, $i=90^\circ$, and $\Omega=-\pi/4$. Shown are points obtained by numerically integrating the SA equations of motion, equations~(\ref{eq:EOM_SA_gen}) (open circles: mean; stars: rms), and the lines correspond to the Newtonian prediction, equations~(\ref{eq:stat_N}) (solid: TO mean; dotted: TO rms). Blue (red) lines correspond to positive (negative) values. The 1PN terms were excluded (included) in the left (right)-hand panels. The vertical yellow dashed lines show equation~(\ref{eq:Qdiva_1PN}), and the vertical black dotted lines show equation~(\ref{eq:Qdiva_1PNb}).}
\label{fig:PN}
\end{figure}

In \F~\ref{fig:PN}, we plot the mean and rms scalar eccentricity changes as a function of $Q/a$ (keeping $a$ fixed while varying $Q$) for a system with $m_1=m_2=\mper=40\,\msun$, $a=1\,\au$, $e=0.999$, $i=90^\circ$, and $\Omega=-\pi/4$. Shown are points each based on 20 integrations with different $\omega_{\mathrm{i}}$, obtained by numerically integrating the SA equations of motion, equations~(\ref{eq:EOM_SA_gen}), and the lines correspond to the Newtonian prediction, equations~(\ref{eq:stat_N}). The 1PN terms were excluded (included) in the left (right)-hand panels. In the numerical integrations, a finite range of $\theta$ should be specified; we integrate from $\theta=-0.95\Leper$ to $\theta=0.95\Leper$. We remark that the numerical results are converged with respect to the integration range. 

In the Newtonian case, the numerically-obtained SA results agree well with the analytic results. Some deviation for the rms values in particular can be seen at very small $Q/a$, for which higher-order terms in $\epssa$ become important. When the 1PN terms are included, the scalar eccentricity changes are consistent with the Newtonian results at small $Q/a$, in which case the rate of Newtonian perturbation is relatively strong compared to the rate of 1PN precession, i.e., 1PN terms are not important. As $Q/a$ increases, the rms eccentricity change in particular shows a drop with respect to the Newtonian case. Also, the mean starts to fluctuate, and change sign. This can be attributed to quenching of the Newtonian torques due to rapid 1PN precession.

The value of $Q/a$ for which 1PN terms start to become important can be estimated by equating the time-scale associated with the Newtonian terms with the 1PN precession time-scale. The former can be estimated by 
\begin{align}
t_{\mathrm{TB}} \sim n_\bin \frac{\mper}{m} \left ( \frac{a}{Q} \right )^3.
\end{align}
Equating the latter to the 1PN time-scale implied by equation~(\ref{eq:1PN_t}), we find for the critical $Q/a$ for 1PN terms to become important
\begin{align}
\label{eq:Qdiva_1PN}
\left (\frac{Q}{a} \right )_{\mathrm{crit}} \sim \left [ \frac{1}{3} \left (1-e^2 \right ) \frac{a}{\rg}\frac{\mper}{m} \right ]^{1/3} \simeq 7.5,
\end{align}
where the numerical value assumes the parameters in \F~\ref{fig:PN}, and which is shown in the latter figure with the vertical yellow dashed lines. It can be seen that the value of $Q/a$ for which 1PN terms start to decrease the typical eccentricity change can be reasonably estimated with this equation. 

Alternatively, the regime in which the 1PN terms dominate can be estimated by equating the time-scale of the passage of the perturber to the 1PN precession time-scale. The former can be estimated as $t_{\mathrm{fly}} \sim \sqrt{Q^3/[G(m+\mper)]}$. This gives
\begin{align}
\label{eq:Qdiva_1PNb}
\left (\frac{Q}{a} \right )_{\mathrm{crit,2}} \sim \left [ \frac{1}{3} \left (1-e^2 \right ) \frac{a}{\rg}\frac{m+\mper}{m} \right ]^{2/3} \simeq 117,
\end{align}
where the numerical value again assumes the parameters in \F~\ref{fig:PN}, and which is shown with the vertical black dotted lines. Approximately, $(Q/a)_{\mathrm{crit,2}} \sim (Q/a)_{\mathrm{crit}}^2$. Equation~(\ref{eq:Qdiva_1PNb}) gives a less conservative estimate for the value of $Q/a$ for which the 1PN terms dominate; the rms $\Delta e$ is already significantly below the Newtonian value at $Q/a =  (Q/a)_{\mathrm{crit,2}}$.

\section{Steady-state distribution}
\label{sect:steady_state}
In \pI~and in the previous sections, we considered in detail the effects of a single encounter on a binary. However, in dense stellar systems such as globular clusters, a typical binary will encounter a large number of perturbers. Although not all encounters will be of the secular type, it is still of theoretical interest to determine the steady-state distribution of encounters in the secular regime. More specifically, encounters are typically thought to give rise to equipartition, i.e., to a uniform distribution in phase space in terms of angular momentum. In other words, in this case, the distribution function $f(\Ec,\Lc)$  is independent of $\Lc$, where $\Ec$ is the orbital energy $\Ec$ and $\Lc$ is the angular momentum (note that $\Lc \propto \j$). The probability density function (PDF) in energy and angular space, $N(\Ec,\Lc)$, is related to the distribution function according to (e.g., \citealt{2013degn.book.....M})
\begin{align}
N(\Ec,\Lc) = 8 \pi^2 \Lc f(\Ec,\Lc) P(\Ec,\Lc),
\end{align}
where $P(\Ec,\Lc)$ is the orbital period. Therefore, if $f(\Ec,\Lc)$ is independent of $\Lc$ (and if $P$ is independent of $\Lc$, which is the case for a binary), then the number distribution in angular momentum is $N(\Lc) \propto \Lc$, which implies a (normalized) PDF in eccentricity
\begin{align}
\label{eq:thermal}
N(e) = 2e.
\end{align}
The well-known distribution in equation~(\ref{eq:thermal}) is also referred to as a thermal distribution \citep{1919MNRAS..79..408J,ambartsumian1937,1975MNRAS.173..729H}. However, it is not a priori clear if a uniform distribution in phase space is reached, and whether strong encounters or more distant encounters contribute to the steady state. Therefore, it is relevant to consider the steady-state distribution arising from secular encounters, which we address here using our analytic expressions for the scalar eccentricity change. Since the calculations are already lengthy when the FO and SO terms are included for parabolic encounters, we will here focus on these simplified cases. We do not include octupole-order terms, nor PN terms, or allow for hyperbolic encounters. We carry out the calculations with only the FO terms included, and with the FO and SO terms both included, to illustrate the importance of adding the SO terms in the (purely Newtonian) steady-state calculations. 

To determine the angular-momentum steady-state distribution, we consider the Fokker-Planck equation in angular-momentum space, which is given by (e.g., \citealt[Section 5.1.1]{2013degn.book.....M})
\begin{align}
\frac{\partial N(\Rt,t)}{\partial t} = - \frac{\partial}{\partial \Rt} [ N(\Rt,t)\Do ] + \frac{1}{2} \frac{\partial^2}{\partial \Rt^2} \left [ N(\Rt,t) \Dt \right ].
\end{align}
Here, $\Rt \equiv \j^2 = 1-e^2$ is the squared normalized angular momentum\footnote{$\Rt$ here is not to be confused with the ratio of the angular speed of the perturber to the binary's mean motion, as defined in equation~(1) of \pI.}, and the brackets denote diffusion coefficients. The steady state is defined by $\partial N(\Rt,t)/ \partial t = 0$, implying
\begin{align}
- N(\Rt,t) \Do + \frac{1}{2} \frac{\partial}{\partial \Rt} \left [ N(\Rt,t) \Dt \right ] = C,
\end{align}
where $C$ is a constant. Here, we consider `zero-flux' solutions, i.e., $C=0$. In this case, it is straightforward to show that the steady-state solution is (e.g., \citealt[Appendix D]{2014MNRAS.443..355H})
\begin{align}
\label{eq:gen_steady_state}
N(\Rt) \propto \exp \left [ \int \mathrm{d} \Rt \frac{2 \Do - \Dt'}{\Dt} \right ],
\end{align}
where the prime denotes the derivative with respect to $\Rt$. 

We compute the diffusion coefficients in $\Rt$ from our analytic expressions for $\Delta e$. Generally, a diffusion coefficient $\langle x \rangle$ in some quantity $x$ measures the change of $x$ per unit of time. In our case, the unit of time corresponds to a single passage of a perturber; also, note that the time unit drops out in the integrand in equation~(\ref{eq:gen_steady_state}). From the definition $\Rt \equiv 1-e^2$, $\Delta \Rt = -2e \Delta e - (\Delta e)^2$, where we retain the second-order term in $\Delta e$. Subsequently, we compute the diffusion coefficients by averaging over all angles of the binary orbit relative to the perturber assuming isotropic orientations, i.e.,
\begin{subequations}
\begin{align}
\Do &= \frac{1}{2\pi} \frac{1}{2\pi} \frac{1}{2} \int_{0}^{2\pi} \mathrm{d} \omega \int_{0}^{2\pi} \mathrm{d} \Omega \int_{0}^{\pi} \mathrm{d} i \, \sin i \left [  -2e \Delta e - (\Delta e)^2 \right ]; \\
\Dt &= \frac{1}{2\pi} \frac{1}{2\pi} \frac{1}{2} \int_{0}^{2\pi} \mathrm{d} \omega \int_{0}^{2\pi} \mathrm{d} \Omega \int_{0}^{\pi} \mathrm{d} i \, \sin i \left [  -2e \Delta e - (\Delta e)^2 \right ]^2.
\end{align}
\end{subequations}
In principle, when computing the diffusion coefficients, one should also average over all perturber properties, such as $\mper$, $\eper$, and impact parameters $b$. Here, we make the simplifying assumption of a single-mass perturber population, and parabolic orbits ($\eper=1$). Also, we assume for simplicity that the octupole-order terms are negligible ($\epsoct=0$). With regards to the impact parameter $b$, we take two approaches. 
\begin{itemize}[leftmargin=5mm]
\item In the first approach (1), we consider $b$ to be fixed and therefore do not average over it, meaning that the results will be valid for a given impact parameter. 
\item In the second approach (2), we also integrate the diffusion coefficients over impact parameters assuming a distribution $N(b) \propto b$. Since we consider fixed $\mper$ and $\eper=1$ (such that $b=Q$, where $Q$ is the perturber's periapsis distance), this implies a (normalized) distribution of perturber strengths 
\begin{align}
N(\epssa) = - \frac{4}{3} \frac{\epssa^{-7/3}}{\epssat^{-4/3} - \epssao^{-4/3}},
\end{align}
where $\epssao$ and $\epssat$ are the values of $\epssa$ corresponding to the strongest (weakest) perturber (i.e., smallest and largest $Q$, respectively). 
\end{itemize}
Below, we state the results separately based on the expressions with the FO terms taken into account only, and also with both the FO and SO terms included. Due to their complexity, we do not include the TO terms here. Moreover, we find numerically that the addition of the TO terms does not significantly affect the steady-state distribution (see \S~\ref{sect:steady_state:MC} below).

\begin{figure}
\center
\includegraphics[scale = 0.46, trim = 8mm 0mm 8mm 0mm]{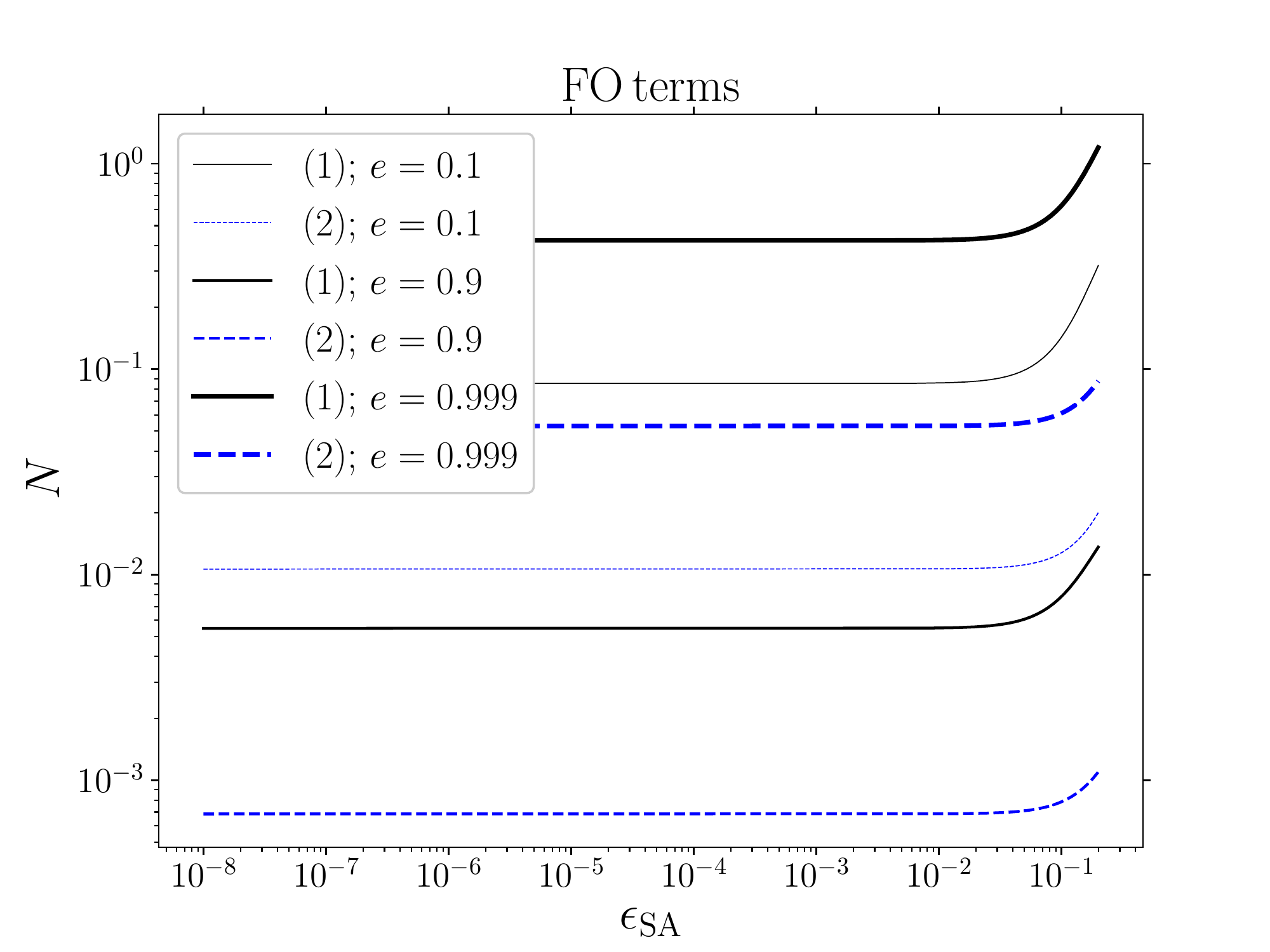}
\includegraphics[scale = 0.46, trim = 8mm 0mm 8mm 0mm]{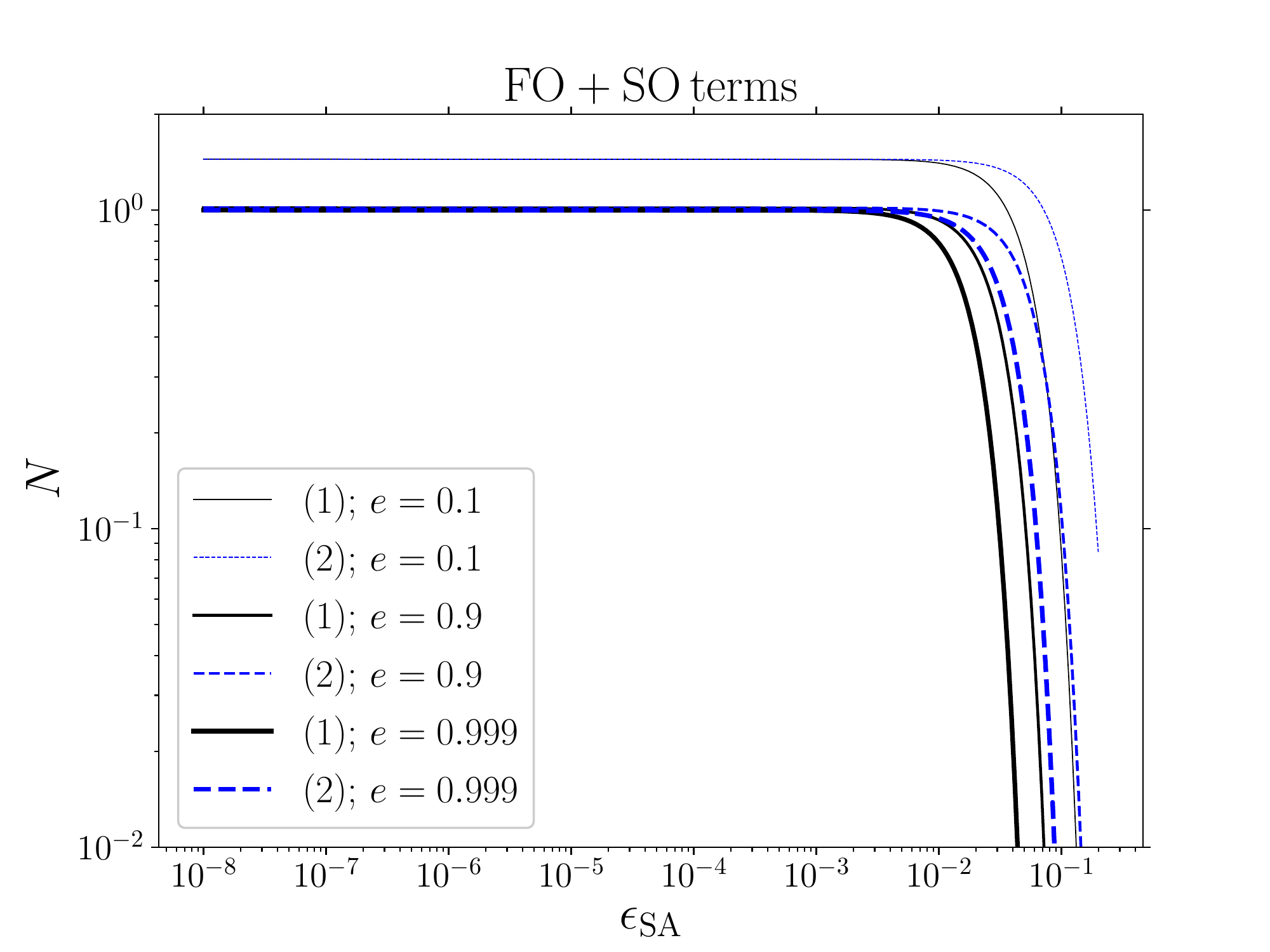}
\caption{ Steady-state distributions for fixed $e$ as a function of $\epssa$ (in the case of approach 1), or as a function of $\epssao$ (in the case of approach 2). Left (right)-hand panels correspond to the distributions with the FO (FO+SO) terms. In each panel, black solid lines correspond to approach (1), whereas blue dashed lines correspond to approach (2). Different values for the eccentricity are assumed, indicated with different line widths ($e$ increases with increasing line width; refer to the legends). }
\label{fig:steady_state_eps}
\end{figure}

\subsection{Steady-state distribution based on FO terms}
\label{sect:steady_state:FO}
In approach (1) (considering $\epssa$ to be fixed) and including the FO terms in $\epssa$ only, the diffusion coefficients are given by (assuming parabolic orbits, and $\epsoct=0$)
\begin{subequations}
\label{eq:DFC_FO}
\begin{align}
\Do &= -\frac{15}{4} \pi^2 \epssa^2 \Rt (1-\Rt); \\
\Dt &= \frac{15}{112} \pi^2 \epssa^2 (1-\Rt)^2 \left(112 + 225\pi^2 \Rt \epssa^2 \right).
\end{align}
\end{subequations}
Applying equation~(\ref{eq:gen_steady_state}), the implied zero-flux steady-state distribution in terms of the eccentricity is given by
\begin{align}
\label{eq:steady_state_FO_1}
N(e) \propto \frac{1}{1-e^2} e^{\frac{112}{225 \pi ^2 \epssa^2+112}-3} \left(112+225 \pi ^2
   \left(1-e^2\right) \epssa^2\right)^{-\frac{56}{225 \pi ^2  \epssa^2+112}-1}
\end{align}
The distribution in equation~(\ref{eq:steady_state_FO_1}) is nearly independent of $\epssa$, unless $\epssa \gtrsim 10^{-2}$. To illustrate, we show with black solid lines in the left-hand panel of \F~\ref{fig:steady_state_eps} the distributions (modulo a normalization factor) as a function of $\epssa$. In the limit of $\epssa\rightarrow0$, equation~(\ref{eq:steady_state_FO_1}) reduces to the simple expression
\begin{align}
\label{eq:steady_state_FO_1_zero}
N(e) \propto \frac{1}{e^2(1-e^2)}.
\end{align}
In approach (2), when making the assumption $\epssat \ll \epssao$, the implied eccentricity distribution is
\begin{align}
\label{eq:steady_state_FO_2}
N(e) \propto \frac{1}{1-e^2} e^{\frac{448}{225 \pi ^2 \epssao^2+448}-3} \left(448+225 \pi ^2
   \left(1-e^2\right) \epssao^2\right)^{-\frac{224}{225 \pi ^2
   \epssao^2+448}-1}.
\end{align}
This distribution is again weakly dependent on $\epssao$, which is illustrated with the blue dashed lines in the left-hand panel of \F~\ref{fig:steady_state_eps}. In the limit of $\epssao\rightarrow0$, equation~(\ref{eq:steady_state_FO_2}) reduces to the same expression as in approach (1), i.e., equation~(\ref{eq:steady_state_FO_1_zero}).

\subsection{Steady-state distribution based on FO and SO terms}
\label{sect:steady_state:SO}
When the SO terms are included, the diffusion coefficients in approach (1) are given by
\begin{subequations}
\label{eq:DFC_SO}
\begin{align}
&\Do = -\frac{3}{1120} \pi ^2 (1-\Rt) \epssa^2 \left(3 \left(625+8158 \pi ^2\right) \Rt^2
   \epssa^2+\Rt \left(6776-24900 \pi ^2 \epssa^2\right)+100
   \left(75 \pi ^2 \epssa^2-28\right)\right); \\
\nonumber &\Dt = \frac{3}{174254080} \pi ^2 (1-\Rt)^2 \epssa^2 \Biggl [ 918 \pi ^2 \left(3125000+113795625
   \pi ^2+434516724 \pi ^4\right) \Rt^4 \epssa^6 \\ 
\nonumber   &\quad -3240 \pi ^2 \Rt^3
   \left(251074980 \pi ^4 \epssa^2+17 \pi ^2 \left(1771875
   \epssa^2-5501464\right)-9116250\right) \epssa^4+3 \Rt^2
   \left(225664380000 \pi ^6 \epssa^4 \right. \\
\nonumber   &\quad \left. +11475 \pi ^4 \left(828125
   \epssa^2-11198592\right) \epssa^2-3536 \pi ^2
   \left(1078125 \epssa^2-4188008\right)+388960000\right) \epssa^2 \\
   &\quad -1600 \Rt \left(167214375 \pi ^6 \epssa^6-108438750 \pi ^4
   \epssa^4+18436704 \pi ^2 \epssa^2-544544\right)+120000
   \pi ^2 \left(354375 \pi ^4 \epssa^4-229500 \pi ^2 \epssa^2+38896\right) \epssa^2\Biggl ].
   \end{align}
\end{subequations}
The general steady-state distribution implied by equations~(\ref{eq:DFC_SO}) is excessively complicated. Here, we simplify the expression by expanding $N(e)$ in $\epssa$, assuming $\epssa\ll 1$. To order $\epssa^2$, the result is
\begin{align}
\label{eq:steady_state_SO_1}
N(e) \propto e^{-\frac{4}{25} -\frac{3 \left(57500+1372461 \pi ^2\right) \epssa^2}{70000}} \left(1-e^2\right)^{\frac{543 \pi ^2 \epssa^2}{14}} \exp \left[-\frac{3 \epssa^2
   \left(182500+7730161 \pi ^2\right) \left(1-e^2\right)}{140000}\right].
\end{align}
Similarly to the FO result, this distribution is insensitive to $\epssa$, unless $\epssa \gtrsim 10^{-2}$ (see the right-hand panel of \F~\ref{fig:steady_state_eps}). Taking the limit $\epssa\rightarrow 0$, the (normalized) distribution simplifies to
\begin{align}
\label{eq:steady_state_SO_1_zero}
N(e) = \frac{21}{25} e^{-4/25}.
\end{align}
In approach (2) and assuming $\epssat\ll \epssao$, the steady-state distribution is
\begin{align}
N(e) \propto e^{-\frac{4}{25} -\frac{3 \left(57500+1372461 \pi ^2\right) \epssao^2}{280000}} \left(1-e^2\right)^{\frac{543 \pi ^2 \epssao^2}{56}} \exp \left [- \frac{3 \epssao^2\left(182500+7730161 \pi ^2\right) \left(1-e^2\right)}{560000} \right ]
\end{align}
Similarly to the FO case, this reduces to equation~(\ref{eq:steady_state_SO_1_zero}) in the limit $\epssao\rightarrow0$.

\subsection{Monte Carlo results}
\label{sect:steady_state:MC}
To verify the distributions derived above, we perform numerical Monte Carlo (MC) experiments in which perturbers are continuously sampled assuming an isotropic orientation relative to the binary, and the effect on the binary eccentricity is computed for each encounter using our analytic FO and SO expressions (we also implemented the TO terms, but their inclusion does not significantly affect the steady-state distribution). Similarly to the analytic expressions, we assume a single perturber mass and parabolic encounter orbits. The masses are set to $m_1=m_2=\mper$, such that $\epsoct=0$. The inner and outer boundaries in terms of $Q$ in the MC experiments are set to be $2\,a$ and $80\,a$, respectively; note that the resulting steady-state distribution is insensitive to these values. 

\begin{figure}
\center
\includegraphics[scale = 0.46, trim = 8mm 0mm 8mm 0mm]{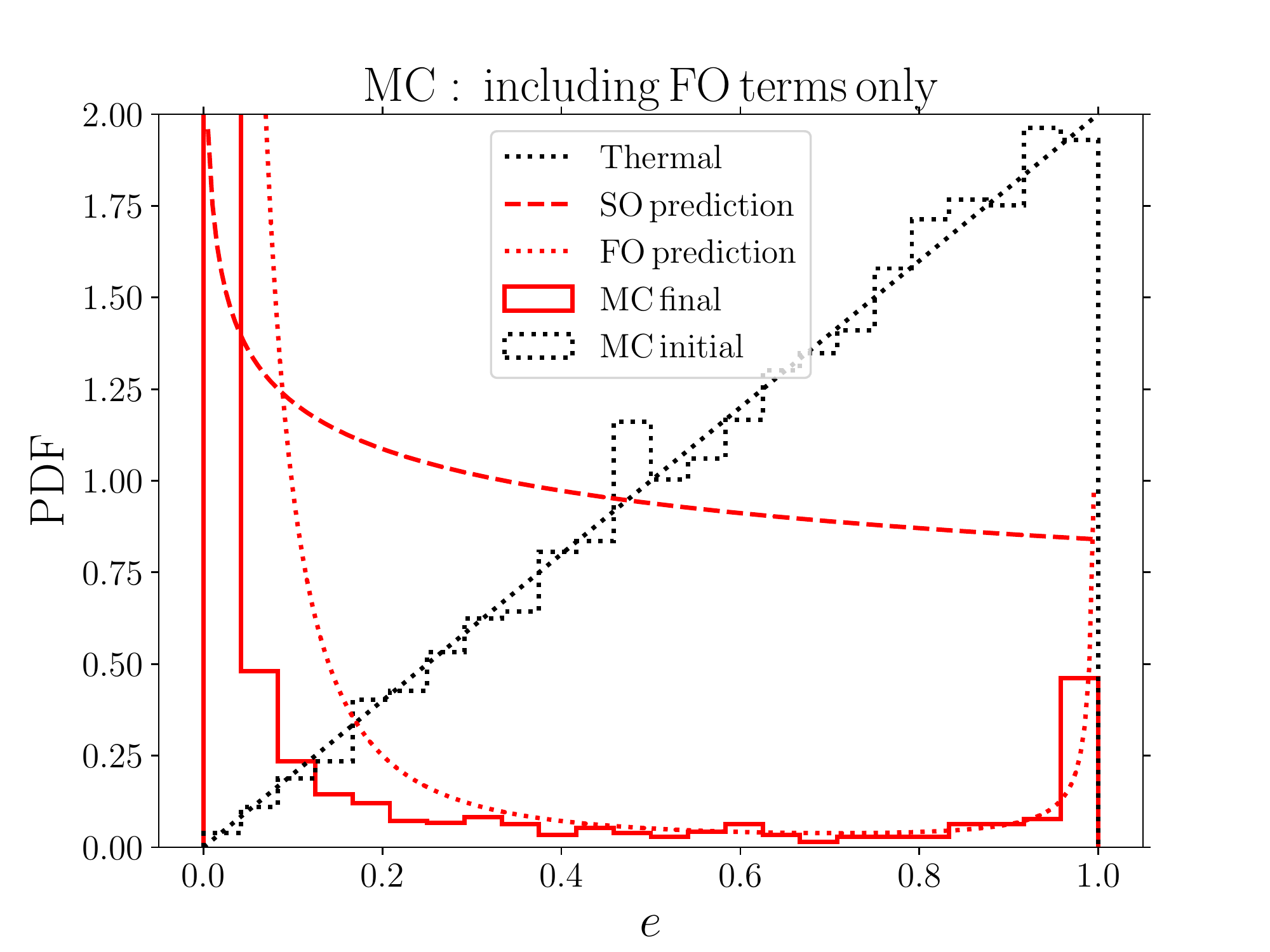}
\includegraphics[scale = 0.46, trim = 8mm 0mm 8mm 0mm]{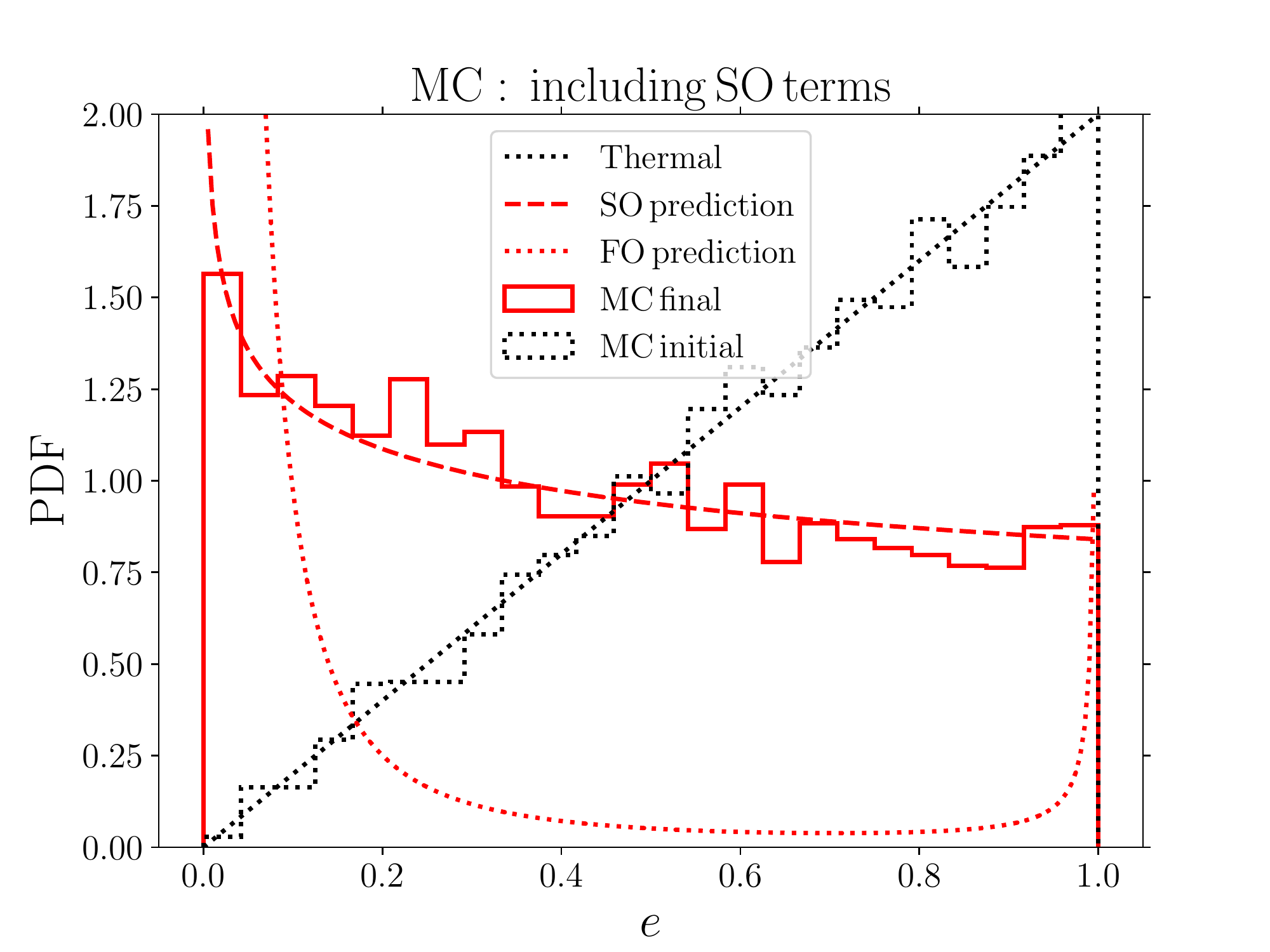}
\caption{ Steady-state eccentricity distributions from the MC experiments in which the FO terms (left-hand panel) and FO and SO terms (right-hand panel) were included. The initial eccentricity distribution of the binary (black dotted lines) is assumed to be thermal; note that the final distributions are independent of the initial distributions. The final eccentricity distributions are shown with the solid red lines. The FO and SO analytic predictions are shown in each panel with red dashed and dotted lines, respectively.  }
\label{fig:steady_state}
\end{figure}

The results are shown in \F~\ref{fig:steady_state}. In the left (right)-hand panels, we show results from the MC experiments in which the FO terms (FO and SO terms) were included. We set the initial eccentricity distribution of the binary to be thermal, i.e., $N(e)=2e$, but we note that the final distributions are independent of the initial distributions (e.g., an initially uniform or delta function distribution). The FO and SO analytic predictions (assuming $\epssa\rightarrow 0$ and $\epssao\rightarrow 0$) are shown in each panel with red dashed and dotted lines, respectively. 

The analytic distributions are in agreement with the numerical results. In particular, with the FO and SO terms, the steady-state distribution is fairly flat, and well described by $N(e)\propto e^{-4/25}$ as derived above. Note that, in the case of the FO terms only, the analytic distribution, equation~(\ref{eq:steady_state_FO_1_zero}), diverges when integrating over eccentricities from 0 to 1; therefore, we normalized between two boundary eccentricities, $e_1$ and $e_2$, and the normalization depends on the precise choices of these values (in \F~\ref{fig:steady_state}, $e_1=0.01$, and $e_2=0.9999$).

\section{Discussion}
\label{sect:discussion}
\subsection{High-order terms}
\label{sect:discussion:higher}
Having presented here higher-order terms compared to \pI, a natural question is whether even higher-order terms can be important. In principle, fourth-order terms in $\epssa$ can be derived using the same procedures. However, as hinted at in Table~\ref{table:N_terms}, the associated number of terms is so large that analytic expressions, although explicit, can no longer be considered to be useful. Moreover, for the fourth-order terms to be important, $\epssa$ must be close to unity, and the secular approximation in this case becomes questionable. 

Higher-order terms could also be derived in the expansion of $r/R$ (the next-order terms in the Hamiltonian are $\propto [r/R]^4$, and are known as the hexadecupole-order terms). This has been done for bound triples (e.g., \citealt{2010A&A...522A..60L,2015MNRAS.452.3610A,2016CeMDA.124...73C,2018MNRAS.481.4602L}), and higher-order multiplicity systems \citep{2015MNRAS.449.4221H,2016MNRAS.459.2827H}. However, in most situations, these higher-order terms do not significantly affect the dynamics. Moreover, similarly to the above, higher-order expansion orders are only important for relatively large $r/R$, and in this case the secular approximation may no longer be adequate.

\subsection{Steady-state distribution}
\label{sect:discussion:steady}
In \S~\ref{sect:steady_state}, we determined the steady-state eccentricity distribution based on our analytic expressions, and verified that our analytic results based on the Fokker-Planck equation are consistent with numerical Monte Carlo experiments. We found that the distributions based on the FO and the FO+SO terms are different from a thermal distribution. The natural question to ask is why this is the case. As discussed in \S~\ref{sect:steady_state}, a thermal distribution results from equipartition in angular momentum. The latter argument does not make a distinction between strong and weak encounters, whereas in our analysis we considered weak (i.e., secular) encounters only. Moreover, in our treatment of the scalar eccentricity change, the perturber is not allowed to exchange angular momentum (nor energy) with the binary -- the binary is perturbed by the third body, but the third body is not affected by the binary. This `test particle' approach may be another reason for a steady state that is different from thermal. 

The above results indicate that the more distant encounters give rise to a different steady-state distribution compared to `all' encounters, which are also allowed to exchange angular momentum (and energy) with the binary. Of course, the latter situation applies to real clusters. Nevertheless, it is instructive to understand the individual contributions from `weak' and `strong' encounters, as the `weak' type of encounters are typically ignored in studies of binaries in clusters (e.g., \citealt{2018PhRvD..97j3014S,2018MNRAS.481.5445S,2019PhRvD..99f3006S}).

\section{Conclusions}
\label{sect:conclusions}
In this paper, we extended previous work in which we analytically calculated the secular effects of a perturber moving on a hyperbolic or parabolic orbit on a binary system. In this approximation, the binary mean motion is much faster than the perturber's motion, and it is justified to average the equations of motion over the binary. The main results are given below.

\medskip \noindent 1. Specifying to parabolic encounters, we derived expressions for the changes of the eccentricity and angular-momentum vectors of the binary to third order (TO) in the perturbation parameter $\epssa$ (cf. equation~\ref{eq:epssa}), and including terms up to and including octupole order in the expansion of the ratio $r/R$ of the binary separation $r$ to the perturber separation $R$. The latter appear in particular for relatively close encounters, and unequal binary component masses ($m_1\neq m_2$). Our methodology is based on Fourier expansions of the equations of motion and the eccentricity and angular-momentum vectors. Starting from the second-order (SO) in $\epssa$, this technique allows to incorporate effects associated with the changing eccentricity and angular-momentum vectors as the perturber passes the binary. The results are summarized in \S~\ref{sect:TO:results}, where a link is given to a publicly-available \textsc{Python} code which can be used to quickly evaluate the expressions. In \S~\ref{sect:TO:test_ex}, we demonstrated the correctness of the TO and octupole expressions, and illustrated the behaviour of secular encounters for parts of the parameter space. 

\medskip \noindent 2. We considered the effects of the lowest-order post-Newtonian (PN) terms, which (after averaging over the binary's orbital phase) give rise to precession of the binary's argument of periapsis. We were unable to find an analytic expression for the scalar eccentricity changes with the addition of the 1PN terms using the methodology of Fourier expansions, due to a divergence occurring in the integrated equations of motion associated with the 1PN terms. Physically, the latter corresponds to the infinite precession of the binary when considering an infinite time span, irrespective of the binary's properties (i.e., for any $a$, $e$, and $m_1$ and $m_2$). Nevertheless, we showed a numerical example of the quenching of the Newtonian perturbation on the binary when the 1PN are included, and the strength of the perturbation is weak (such that the 1PN terms dominate). We found that the transition between the two regimes (1PN terms not being important, and the 1PN terms quenching the perturbation) can be estimated using equations~(\ref{eq:Qdiva_1PN}) or (\ref{eq:Qdiva_1PNb}).

\medskip \noindent 3. We determined the steady-state binary eccentricity distribution in response to the cumulative effect of secular encounters by computing the associated angular-momentum diffusion coefficients, and applying the Fokker-Planck equation. For simplicity, we restricted to the steady-state distribution according to both our first-order (FO) and second-order (SO) scalar eccentricity change expressions for parabolic encounters, and setting the octupole-order terms to zero ($m_1=m_2$). We found that the steady-state distributions in both the FO and SO cases are weakly dependent on the perturber impact parameter, unless $\epssa \gtrsim 10^{-2}$. In the limit $\epssa \rightarrow 0$, we found that the steady-state eccentricity distribution is $N(e) \propto [e^2(1-e^2)]^{-1}$, whereas for the SO expression, we found $N(e) \propto e^{-4/25}$. The FO expression is strongly biased to circular and highly eccentric orbits. The SO distribution, on the other hand, is much flatter, with some preference for circular orbits, and no high-eccentricity tail. We verified these analytic expressions numerically by comparing them to results from Monte Carlo experiments.

\section*{Acknowledgements}
We thank Scott Tremaine for stimulating discussions and comments on the manuscript, and the referee, Nathan Leigh, for a very helpful report. A.S.H. gratefully acknowledges support from the Institute for Advanced Study, and the Martin A. and Helen Chooljian Membership. J.S. acknowledges support from the Lyman Spitzer Fellowship.

\bibliographystyle{mnras}
\bibliography{literature}

\appendix


\section{Explicit expressions}
\label{app:fgh}
Here, we give the explicit vector functions that appear in $\Delta \ve{e}$ and $\Delta \ve{\j}$ (equation~\ref{eq:Delta_e_Delta_j}). Given their excessive length, we do not show the expressions for the functions associated with the octupole-order terms at TO in $\epssa$. 

The functions associated with the vector eccentricity change are given by
\begin{dmath*}[style={\scriptsize}]
\feI = \left\{-\frac{3}{2} \pi  (e_y \j_z+3 e_z \j_y),\frac{3}{2} \pi   (e_x \j_z+3 e_z \j_x),3 \pi  (e_y \j_x-e_x \j_y)\right\}; 
\end{dmath*}
\begin{dmath*}[style={\scriptsize}]
\feII = \left\{-\frac{15}{16} \pi   (-3 e_x e_y \j_z-27 e_x e_z \j_y+5 e_y e_z \j_x+5 \j_x \j_y \j_z),-\frac{15}{32} \pi   \left(9 e_x^2 \j_z+34 e_x e_z \j_x+3 e_y^2 \j_z-10 e_y e_z \j_y-\j_z \left(32 e_z^2+15 \j_x^2+5 \j_y^2-4\right)\right),\frac{15}{32} \pi   \left(19 e_x^2 \j_y-26 e_x e_y \j_x-\j_y \left(7 e_y^2+32 e_z^2+5 \j_x^2+5 \j_y^2-4\right)\right)\right\};
\end{dmath*}
\begin{dmath*}[style={\scriptsize}]
\geI = \left\{-\frac{3}{16} \pi   \left(-75 e_x^2 e_y+6 \pi  e_x \left(15 e_z^2-6 \j_y^2+\j_z^2\right)-50 e_y^3+e_y \left(-50 e_z^2-5 \j_x^2+36 \pi  \j_x \j_y+50 \left(\j_y^2+2 \j_z^2\right)\right)-50 e_z \j_y \j_z\right),-\frac{3}{16} \pi   \left(75 e_x^3+e_x \left(50 e_y^2+5 \j_x^2+36 \pi  \j_x \j_y-150 \j_z^2\right)+e_y \left(90 \pi  e_z^2-36 \pi  \j_x^2-50 \j_x \j_y+6 \pi  \j_z^2\right)-10 e_z \j_x \j_z\right),-\frac{3}{8} \pi   \left(30 \pi  e_x^2 e_z+e_x (25 e_y e_z-12 \pi  \j_x \j_z+75 \j_y \j_z)+e_z \left(30 \pi  e_y^2-18 \pi  \j_x^2+25 \j_x \j_y-18 \pi  \j_y^2\right)-3 e_y \j_z (15 \j_x+4 \pi  \j_y)\right)\right\};
\end{dmath*}
\begin{dmath*}[style={\scriptsize}]
\geII = \left\{-\frac{15}{512} \pi   \left(2541 e_x^3 e_y+e_x^2 \left(36 \pi  \left(e_y^2-163 e_z^2+55 \j_y^2-5 \j_z^2\right)-847 \j_x \j_y\right)+e_x \left(2037 e_y^3-e_y \left(294 e_z^2+637 \j_x^2+2592 \pi  \j_x \j_y+3969 \j_y^2+4074 \j_z^2-420\right)+24 e_z \j_z (76 \pi  \j_x-77 \j_y)\right)+60 \pi  \j_x^2 \left(3 e_y^2+15 e_z^2-3 \j_y^2+\j_z^2\right)+7 \j_x \left(\j_y \left(163 e_y^2-482 e_z^2-62 \j_z^2+20\right)-264 e_y e_z \j_z+\j_y^3\right)+12 \pi  \left(3 e_y^4+e_y^2 \left(-91 e_z^2-36 \j_y^2+3 \j_z^2+4\right)+72 e_y e_z \j_y \j_z+96 e_z^4+e_z^2 \left(81 \j_y^2+32 \j_z^2-12\right)-15 \j_y^4+15 \j_y^2 \j_z^2+12 \j_y^2-4 \j_z^2\right)-49 \j_x^3 \j_y\right),\\
\frac{15}{512} \pi   \left(2541 e_x^4+36 \pi  e_x^3 e_y+e_x^2 \left(2037 e_y^2-2513 e_z^2-1484 \j_x^2+1380 \pi  \j_x \j_y-903 \j_y^2-7623 \j_z^2+420\right)+4 e_x \left(9 \pi  e_y^3+e_y \left(3 \pi  \left(398 e_z^2-35 \j_y^2+18 \j_z^2+4\right)-453 \pi  \j_x^2-658 \j_x \j_y\right)-2 e_z \j_z (469 \j_x+36 \pi  \j_y)\right)-e_z^2 \left(707 e_y^2+3213 \j_x^2-72 \pi  \j_x \j_y+7 \left(\j_y^2-480 \j_z^2+20\right)\right)+707 e_y^2 \j_x^2-12 \pi  e_y^2 \j_x \j_y-2037 e_y^2 \j_z^2+56 e_y e_z \j_z (47 \j_y-12 \pi  \j_x)+1120 e_z^4-49 \j_x^4-180 \pi  \j_x^3 \j_y+7 \j_x^2 \j_y^2+637 \j_x^2 \j_z^2+140 \j_x^2-180 \pi  \j_x \j_y^3+120 \pi  \j_x \j_y \j_z^2+144 \pi  \j_x \j_y+903 \j_y^2 \j_z^2-420 \j_z^2\right),\\
\frac{15}{512} \pi   \left(2892 \pi  e_x^3 e_z-e_y \left(e_z \left(-2219 e_x^2+1561 \j_x^2+264 \pi  \j_x \j_y+3059 \j_y^2-140\right)+2 e_x \j_z (2891 \j_x+780 \pi  \j_y)+1120 e_z^3\right)+3 e_x^2 \j_z (3143 \j_y-492 \pi  \j_x)+e_y^2 (2892 \pi  e_x e_z+84 \pi  \j_x \j_z+2471 \j_y \j_z)-2 e_x e_z \left(6 \pi  \left(224 e_z^2+179 \j_y^2-28\right)+1206 \pi  \j_x^2-679 \j_x \j_y\right)+707 e_y^3 e_z+3 \j_z \left(-7 \j_y \left(160 e_z^2+51 \j_x^2-20\right)+4 \pi  \j_x \left(96 e_z^2+35 \j_x^2-12\right)+140 \pi  \j_x \j_y^2-301 \j_y^3\right)\right)\right\};
\end{dmath*}
\begin{dmath*}[style={\tiny},]
\geIII = \left\{\frac{225}{32768} \pi  \left(13041 e_y^5+288 e_x \pi  e_y^4+\left(31626 e_x^2+36092 e_z^2+2458 \j_x^2-23654 \j_y^2-52164 \j_z^2+640 \j_x \j_y \pi -672\right) e_y^3+4 \left(72 \pi  e_x^3+\left(920 \pi  \j_x^2+1583 \j_y \j_x+8 \left(-527 e_z^2-234 \j_y^2+9 \j_z^2+12\right) \pi \right) e_x+2 e_z \j_z (200 \pi  \j_x+11097 \j_y)\right) e_y^2+\left(20097 e_x^4+2 \left(12502 e_z^2+507 \j_x^2-32777 \j_y^2-31626 \j_z^2-13472 \j_x \j_y \pi +2016\right) e_x^2+48 e_z \j_z (53 \j_x+216 \j_y \pi ) e_x+68096 e_z^4+3825 \j_x^4+1925 \j_y^4-3136 \j_x^2-1570 \j_x^2 \j_y^2+1568 \j_y^2-4916 \j_x^2 \j_z^2-14124 \j_y^2 \j_z^2+1344 \j_z^2 \right ) e_y + \left (4 e_z^2 \left(8067 \j_x^2+400 \j_y \pi  \j_x-19099 \j_y^2-2688 \j_z^2-3248\right)+3200 \j_x \j_y^3 \pi -960 \j_x \j_y \j_z^2 \pi +1600 \j_x^3 \j_y \pi -1280 \j_x \j_y \pi +560\right) e_y+8 e_x^2 e_z \j_z (4472 \pi  \j_x+1167 \j_y)-8 e_z \j_z \left(-3883 \j_y^3+1080 \j_x \pi  \j_y^2+\left(-17024 e_z^2-5767 \j_x^2+2128\right) \j_y+40 \j_x \left(32 e_z^2+15 \j_x^2-4\right) \pi \right)-4 e_x^3 \left(7343 \j_x \j_y+216 \left(73 e_z^2-19 \j_y^2+\j_z^2\right) \pi \right)+4 e_x \left(281 \j_y \j_x^3+40 \left(115 e_z^2-17 \j_y^2-21 \j_z^2\right) \pi  \j_x^2+\j_y \left(-20602 e_z^2+2237 \j_y^2-3802 \j_z^2+672\right) \j_x+8 \left(864 e_z^4+3 \left(243 \j_y^2+32 \j_z^2-36\right) e_z^2-135 \j_y^4-12 \j_z^2+\j_y^2 \left(65 \j_z^2+108\right)\right) \pi \right)\right) ,\\
-\frac{225}{32768} \pi  \left(20097 e_x^5+432 e_y \pi  e_x^4+\left(31626 e_y^2-2884 e_z^2-28358 \j_x^2-8294 \j_y^2-80388 \j_z^2+8896 \j_x \j_y \pi +4032\right) e_x^3+4 \left(144 \pi  e_y^3+\left(-3896 \pi  \j_x^2-10513 \j_y \j_x+8 \left(1363 e_z^2-170 \j_y^2+27 \j_z^2+24\right) \pi \right) e_y+2 e_z \j_z (2697 \j_x+472 \j_y \pi )\right) e_x^2+
\left(13041 e_y^4+2 \left(7126 e_z^2-3209 \j_x^2+3531 \j_y^2-31626 \j_z^2-64 \j_x \j_y \pi -336\right) e_y^2+16 e_z \j_z (6207 \j_y-1256 \j_x \pi ) e_y+57344 e_z^4+4949 \j_x^4+2145 \j_y^4-448 \j_x^2 \right )e_x+\left (14558 \j_x^2 \j_y^2-2464 \j_y^2-2028 \j_x^2 \j_z^2+16588 \j_y^2 \j_z^2-8064 \j_z^2-4 e_z^2 \left(10699 \j_x^2-4272 \j_y \pi  \j_x-2691 \j_y^2-16128 \j_z^2+2912\right)-1920 \j_x \j_y^3 \pi -2880 \j_x \j_y \j_z^2 \pi -320 \j_x^3 \j_y \pi +1536 \j_x \j_y \pi +560\right) e_x
+4 \left(36 \pi  e_y^5-\left(240 \pi  \j_x^2+7679 \j_y \j_x+8 \left(81 e_z^2-5 \j_y^2+9 \j_z^2-12\right) \pi \right) e_y^3+6 e_z \j_z (613 \j_x+312 \j_y \pi ) e_y^2\right )
+4\left (\left(300 \pi  \j_x^4-1795 \j_y \j_x^3+40 \left(15 e_z^2+30 \j_y^2+39 \j_z^2-8\right) \pi  \j_x^2-\j_y \left(20602 e_z^2+55 \j_y^2+3802 \j_z^2-1008\right) \j_x+4 \left(704 e_z^4+6 \left(135 \j_y^2+32 \j_z^2-36\right) e_z^2+125 \j_y^4-24 \j_z^2-10 \j_y^2 \left(7 \j_z^2+12\right)+16\right) \pi \right) e_y\right )+4\left (2 e_z \j_z \left(4219 \j_x^3-40 \j_y \pi  \j_x^2+\left(14336 e_z^2+2407 \j_y^2-1792\right) \j_x+88 \j_y \left(32 e_z^2+5 \j_y^2-4\right) \pi \right)\right)
\right) ,\\
-\frac{225}{2048} \pi  \left(1024 \pi  e_z^5+32 \left(20 \pi  \j_x^2+21 \j_y \j_x+4 \left(e_y^2-11 \j_y^2-2\right) \pi \right) e_z^3+32 e_y \j_z (77 \j_x+24 \j_y \pi ) e_z^2+\left(75 \pi  \j_x^4+519 \j_y \j_x^3-10 \left(17 e_y^2+17 \j_y^2+8\right) \pi  \j_x^2+\j_y \left(4307 e_y^2+573 \j_y^2-84\right) \right ) e_z  
\right ) \\
- \frac{225}{2048 \pi} \left ( \left ( \j_x-\left(21 e_y^4+2 \left(159 \j_y^2+8\right) e_y^2+245 \j_y^4-176 \j_y^2-16\right) \pi \right) e_z+1387 e_x^4 \pi  e_z+e_y \j_z \left(649 \j_x^3-100 \j_y \pi  \j_x^2+\left(-2953 e_y^2+769 \j_y^2-308\right) \j_x+4 \j_y \left(7 e_y^2+55 \j_y^2-24\right) \pi \right)+e_x^3 (1743 e_y e_z+6061 \j_y \j_z-804 \j_x \j_z \pi )\\
+e_x \left(1365 e_z e_y^3+\j_z (28 \pi  \j_x+4021 \j_y) e_y^2+e_z \left(672 e_z^2+1769 \j_x^2-6557 \j_y^2+344 \j_x \j_y \pi -84\right) e_y \right )+e_x \left (\j_z \left(-1573 \j_y^3+860 \j_x \pi  \j_y^2+\left(-6496 e_z^2-1525 \j_x^2+812\right) \j_y+4 \j_x \left(448 e_z^2+135 \j_x^2-56\right) \pi \right)\right)\\
-e_x^2 \left(2944 \pi  e_z^3+\left(1534 \pi  \j_x^2+3047 \j_y \j_x-2 \left(683 e_y^2-1013 \j_y^2+184\right) \pi \right) e_z+e_y \j_z (4777 \j_x+804 \j_y \pi )\right)\right) \right\};
\end{dmath*}
\begin{dmath*}[style={\tiny}]
\heI = \left\{\frac{3}{512} \pi   \left(-1200 \pi  e_x^3 \j_z+15 e_x^2 \left(1361 e_y \j_z+80 \pi  e_z \j_x+\left(369+384 \pi ^2\right) e_z \j_y\right)-6 e_x \left(300 \pi  e_y^2 \j_z+15 e_y e_z \left(\left(64 \pi ^2-109\right) \j_x+160 \pi  \j_y\right)+\j_z \left(120 \pi  \j_x^2+\left(581+192 \pi ^2\right) \j_x \j_y+20 \pi  \j_y^2\right)\right)-7575 e_y^3 \j_z+15 e_y^2 e_z (440 \pi  \j_x+681 \j_y)+e_y \j_z \left(\left(1152 \pi ^2-3353\right) \j_x^2+960 \pi  \j_x \j_y+399 \j_y^2\right)+e_z \left(720 \pi  \j_x^3-313 \j_x^2 \j_y+1080 \pi  \j_x \j_y^2-927 \j_y^3\right)\right),\frac{9}{512} \pi   \left(-4475 e_x^3 \j_z-5 e_x^2 (80 \pi  e_y \j_z+177 e_z \j_x-480 \pi  e_z \j_y)+e_x \left(5655 e_y^2 \j_z+10 e_y e_z \left(\left(65+192 \pi ^2\right) \j_y-200 \pi  \j_x\right)+\j_z \left(1043 \j_x^2-80 \pi  \j_x \j_y-\left(1001+384 \pi ^2\right) \j_y^2\right)\right)-600 \pi  e_y^3 \j_z-5 e_y^2 e_z \left(\left(384 \pi ^2-387\right) \j_x+520 \pi  \j_y\right)+2 e_y \j_z \left(-80 \pi  \j_x^2+\left(192 \pi ^2-371\right) \j_x \j_y+140 \pi  \j_y^2\right)+e_z \left(29 \j_x^3+240 \pi  \j_x^2 \j_y+95 \j_x \j_y^2+360 \pi  \j_y^3\right)\right),-\frac{3}{64} \pi   \left(2400 \pi  e_x^2 e_z \j_z+e_y \left(-420 e_x e_z \j_z+45 \left(21+16 \pi ^2\right) e_z^2 \j_x+\j_z^2 \left(48 \pi ^2 \j_x+7 \j_x-240 \pi  \j_y\right)\right)+15 e_x e_z^2 \left(80 \pi  \j_x+\left(133-48 \pi ^2\right) \j_y\right)+e_x \j_z^2 \left(\left(77-48 \pi ^2\right) \j_y-480 \pi  \j_x\right)+1200 \pi  e_y^2 e_z \j_z-12 e_z \j_x \j_z (20 \pi  \j_x+49 \j_y)\right)\right\};
\end{dmath*}

The functions associated with the vector angular-momentum changes are given by
\begin{dmath*}[style={\scriptsize}]
\fjI = \left\{-\frac{3}{2} \pi   (5 e_y e_z-\j_y \j_z),\frac{3}{2} \pi   (5 e_x e_z-\j_x \j_z),0\right\};
\end{dmath*}
\begin{dmath*}[style={\scriptsize}]
\fjII = \left\{-\frac{75}{16} \pi   (-7 e_x e_y e_z+e_x \j_y \j_z+e_y \j_x \j_z+e_z \j_x \j_y),\frac{15}{32} \pi   \left(e_z \left(-73 e_x^2-3 e_y^2+15 \j_x^2+5 \j_y^2-4\right)+10 \j_z (3 e_x \j_x+e_y \j_y)+32 e_z^3\right),\frac{15}{32} \pi   \left(e_y \left(3 e_x^2-32 e_z^2-5 \j_x^2-15 \j_y^2+4\right)-10 e_x \j_x \j_y+3 e_y^3\right)\right\};
\end{dmath*}
\begin{dmath*}[style={\scriptsize}]
\gjI = \left\{\frac{3}{16} \pi   \left(75 e_x^2 \j_y+60 \pi  e_x e_y \j_y+\j_x \left(5 \j_x \j_y-6 \pi  \left(10 e_y^2+\j_z^2\right)\right)-50 e_y e_z \j_z+10 e_z^2 (5 \j_y-9 \pi  \j_x)\right),\\
-\frac{3}{16} \pi   \left(15 e_x^2 (5 \j_x+4 \pi  \j_y)-10 e_x (6 \pi  e_y \j_x+5 e_y \j_y+15 e_z \j_z)+50 e_y^2 \j_x+90 \pi  e_z^2 \j_y+5 \j_x^3-10 \j_x \j_z^2+6 \pi  \j_y \j_z^2\right),\\
-\frac{15}{8} \pi   (5 e_x e_y \j_z+5 e_x e_z \j_y+5 e_y e_z \j_x+\j_x \j_y \j_z)\right\};
\end{dmath*}
\begin{dmath*}[style={\scriptsize}]
\gjII = \left\{\frac{15}{512} \pi   \left(-2541 e_x^3 \j_y+e_y \left(7 \j_x \left(121 e_x^2+682 e_z^2-63 \j_y^2+62 \j_z^2-20\right)-24 \left(116 \pi  e_x^2 \j_y-322 e_x e_z \j_z+\pi  \j_y \left(41 e_z^2-5 \j_y^2-5 \j_z^2+8\right)\right)+49 \j_x^3\right)-96 \pi  e_x^2 e_z \j_z+3 e_y^2 (1080 \pi  e_x \j_x+343 e_x \j_y-32 \pi  e_z \j_z)+e_x \left(7 \j_y \left(42 e_z^2+91 \j_x^2-258 \j_z^2-60\right)+360 \pi  \j_x \left(11 e_z^2+\j_z^2\right)+120 \pi  \j_x \j_y^2+903 \j_y^3\right)+e_y^3 (456 \pi  \j_y-707 \j_x)-64 e_z \j_z \left(3 \pi  \left(8 e_z^2-1\right)+15 \pi  \j_x^2-7 \j_x \j_y\right)\right),\\
\frac{15}{256} \pi   \left(e_x^3 (847 \j_x+1446 \pi  \j_y)-e_x^2 (3 e_y (558 \pi  \j_x+511 \j_y)+6748 e_z \j_z)+e_x \left(7 \j_x \left(196 e_y^2-359 e_z^2-64 \j_y^2+91 \j_z^2+40\right)-6 \pi  \j_y \left(29 e_y^2-406 e_z^2+35 \j_y^2-50 \j_z^2-28\right)-343 \j_x^3-270 \pi  \j_x^2 \j_y\right)-54 \pi  e_y^3 \j_x-1372 e_y^2 e_z \j_z+e_y \left(3 e_z^2 (12 \pi  \j_x+511 \j_y)+210 \pi  \j_x^3+217 \j_x^2 \j_y+6 \pi  \j_x \left(25 \j_y^2-30 \j_z^2-12\right)-217 \j_y \j_z^2\right)+8 e_z \j_z \left(280 e_z^2+49 \j_x^2-60 \pi  \j_x \j_y+56 \j_y^2-35\right)\right),\\
\frac{15}{512} \pi   \left(36 \pi  e_x^3 \j_z+e_x^2 (3549 e_y \j_z-492 \pi  e_z \j_x+2219 e_z \j_y)+2 e_x \left(18 \pi  e_y^2 \j_z+e_y e_z (60 \pi  \j_y-21 \j_x)+\j_z \left(-6 \pi  \left(32 e_z^2+25 \j_y^2-4\right)+30 \pi  \j_x^2+49 \j_x \j_y\right)\right)+e_z \left(-7 \j_y \left(337 e_y^2+23 \j_x^2-20\right)-36 \pi  \j_x \left(17 e_y^2+5 \j_x^2-4\right)-180 \pi  \j_x \j_y^2+7 \j_y^3\right)+e_y \j_z \left(2037 e_y^2-1071 \j_x^2+360 \pi  \j_x \j_y-469 \j_y^2+420\right)-3360 e_y e_z^2 \j_z-32 e_z^3 (36 \pi  \j_x+35 \j_y)\right)\right\};
\end{dmath*}
\begin{dmath*}[style={\scriptsize}]
\gjIII = \left\{\frac{225}{32768} \pi  \left(2145 \j_y^5+800 \j_x \pi  \j_y^4+22 \left(311 e_y^2+906 e_z^2+255 \j_x^2-390 \j_z^2-112\right) \j_y^3+40 \left(88 \j_x \pi  e_y^2+649 e_z \j_z e_y+4 \j_x \left(37 e_z^2+5 \j_x^2+5 \j_z^2-4\right) \pi \right) \j_y^2+20097 e_x^4 \j_y+\left(-17675 e_y^4+6 \left(126 e_z^2-787 \j_x^2-2354 \j_z^2+560\right) e_y^2-9600 e_z \j_x \j_z \pi  e_y+68096 e_z^4+3825 \j_x^4-3136 \j_x^2-11220 \j_x^2 \j_z^2+4928 \j_z^2+4 e_z^2 \left(8067 \j_x^2-9856 \j_z^2-3248\right)+560\right) \j_y-4 e_x^3 (7343 e_y \j_x-5200 e_y \j_y \pi -400 e_z \j_z \pi )-8 \left(60 \j_x \pi  e_y^4+1337 e_z \j_z e_y^3-20 \j_x \left(-35 e_z^2+5 \j_x^2+13 \j_z^2-4\right) \pi  e_y^2-23 e_z \left(896 e_z^2+285 \j_x^2-112\right) \j_z e_y+20 \j_x \left(32 e_z^2+15 \j_x^2-4\right) \left(e_z^2+\j_z^2\right) \pi \right)-4 e_x \left((2219 \j_x+2080 \j_y \pi ) e_y^3+1840 e_z \j_z \pi  e_y^2-\left(281 \j_x^3+240 \j_y \pi  \j_x^2+\left(-11690 e_z^2+6039 \j_y^2-3802 \j_z^2+672\right) \j_x+80 \j_y \left(93 e_z^2-10 \j_y^2-3 \j_z^2+12\right) \pi \right) e_y\right )-4e_x\left (4 e_z \j_z \left(-940 \pi  \j_x^2+2069 \j_y \j_x+20 \left(-96 e_z^2+7 \j_y^2+12\right) \pi \right)\right)-2 e_x^2 \left(4147 \j_y^3+720 \j_x \pi  \j_y^2+\left(12817 e_y^2-12502 e_z^2-507 \j_x^2-8294 \j_z^2-2016\right) \j_y+4 \left(3700 \j_x \pi  e_y^2+8813 e_z \j_z e_y+60 \j_x \left(55 e_z^2+7 \j_z^2\right) \pi \right)\right)\right) ,\\
-\frac{225}{32768} \pi  \left(3825 \j_x^5+1200 \j_y \pi  \j_x^4+2 \left(1069 e_x^2+4320 e_y \pi  e_x+1229 e_y^2+11982 e_z^2+2805 \j_y^2-7650 \j_z^2-1568\right) \j_x^3+4 \left(-1400 \j_y \pi  e_x^2+(2007 e_y \j_y+11194 e_z \j_z) e_x+40 \left(10 \j_y^3+\left(14 e_y^2+111 e_z^2+15 \j_z^2-8\right) \j_y+54 e_y e_z \j_z\right) \pi \right) \j_x^2-\left(9275 e_x^4+31808 e_y \pi  e_x^3+\left(-22750 e_y^2+8652 e_z^2-654 \j_y^2+2028 \j_z^2-6720\right) e_x^2+64 \left(42 e_y^3+\left(77 e_z^2-170 \j_y^2+45 \j_z^2+56\right) e_y+610 e_z \j_y \j_z\right) \pi  e_x\right )-\left (-13041 e_y^4-57344 e_z^4-2145 \j_y^4+11648 e_z^2-10764 e_z^2 \j_y^2+2464 \j_y^2+50176 e_z^2 \j_z^2+11220 \j_y^2 \j_z^2-6272 \j_z^2+848 e_y e_z \j_y \j_z+e_y^2 \left(-14252 e_z^2-7062 \j_y^2+4916 \j_z^2+672\right)-560\right) \j_x+4 \left(5548 \j_y \pi  e_x^4-7 (2045 e_y \j_y+4126 e_z \j_z) e_x^3-8 \left(230 \j_y^3+\left(242 e_y^2-2227 e_z^2-195 \j_z^2-184\right) \j_y+274 e_y e_z \j_z\right) \pi  e_x^2\right )+ 4\left (\left(-7679 \j_y e_y^3-10906 e_z \j_z e_y^2+\j_y \left(4438 e_z^2-55 \j_y^2-3802 \j_z^2+1008\right) e_y+2 e_z \left(28672 e_z^2+5883 \j_y^2-3584\right) \j_z\right) e_x\right ) \\
+ 4\left (4 \left(25 \j_y^5+10 \left(17 e_y^2+59 e_z^2-5 \j_z^2-4\right) \j_y^3-220 e_y e_z \j_z \j_y^2+\left(-51 e_y^4+\left(986 e_z^2-70 \j_z^2-56\right) e_y^2+8 \left(8 e_z^2-1\right) \left(43 e_z^2-5 \j_z^2-2\right)\right) \j_y+4 e_y e_z \left(3 e_y^2-32 e_z^2+4\right) \j_z\right) \pi \right)\right), \\
-\frac{225}{2048} \pi  \left(27 \j_z \pi  e_x^4+e_z (1743 \j_y-628 \j_x \pi ) e_x^3+\j_z \left(210 \pi  \j_x^2+213 \j_y \j_x-2 \left(192 e_z^2+125 \j_y^2-24\right) \pi \right) e_x^2+e_z \left(573 \j_y^3-20 \j_x \pi  \j_y^2+\left(672 e_z^2-459 \j_x^2-84\right) \j_y-4 \j_x \left(192 e_z^2+5 \j_x^2-24\right) \pi \right) e_x\\
+e_y^3 (1365 e_z \j_x+693 e_x \j_z-628 e_z \j_y \pi )+\j_z \left(75 \pi  \j_x^4+255 \j_y \j_x^3+10 \left(64 e_z^2+15 \j_y^2-8\right) \pi  \j_x^2+3 \j_y \left(224 e_z^2+55 \j_y^2-28\right) \j_x+\left(1024 e_z^4+128 \left(5 \j_y^2-2\right) e_z^2+75 \j_y^4-80 \j_y^2+16\right) \pi \right)\\
+e_y \left(96 (7 \j_x-8 \j_y \pi ) e_z^3-4704 e_x \j_z e_z^2+\left(519 \j_x^3-20 \j_y \pi  \j_x^2-3 \left(273 e_x^2+99 \j_y^2+28\right) \j_x-4 \j_y \left(157 e_x^2+5 \j_y^2-24\right) \pi \right) e_z+e_x \j_z \left(1071 e_x^2-1131 \j_x^2-969 \j_y^2+920 \j_x \j_y \pi +588\right)\right)\\
+e_y^2 \left(54 \j_z \pi  e_x^2-e_z (628 \pi  \j_x+1953 \j_y) e_x+\j_z \left(-250 \pi  \j_x^2+375 \j_y \j_x+6 \left(-64 e_z^2+35 \j_y^2+8\right) \pi \right)\right)+27 e_y^4 \j_z \pi \right) \right\};
\end{dmath*}
\begin{dmath*}[style={\scriptsize}]
\hjI = \left\{\frac{3}{512} \pi   \left(-5 e_x^2 \left(2235 e_y e_z-\left(4733+384 \pi ^2\right) \j_y \j_z\right)+10 e_x \left(60 \pi  e_z \left(5 e_y^2-13 \j_y^2\right)+\left(239-192 \pi ^2\right) e_y \j_x \j_z+3 \left(413+192 \pi ^2\right) e_z \j_x \j_y\right)-3 \left(275 e_y^3 e_z-5 e_y^2 \j_z (23 \j_y-200 \pi  \j_x)+5 e_y e_z \left(\left(384 \pi ^2-119\right) \j_x^2-175 \j_y^2\right)+\j_y \j_z \left(149 \j_x^2-40 \pi  \j_x \j_y+11 \j_y^2\right)\right)\right),\\
\frac{3}{512} \pi   \left(5025 e_x^3 e_z-15 e_x^2 \j_z (1199 \j_x+720 \pi  \j_y)-5 e_x \left(1545 e_y^2 e_z-2 e_y \j_z \left(1080 \pi  \j_x+\left(1105+192 \pi ^2\right) \j_y\right)+3 e_z \left(7 \j_x^2-80 \pi  \j_x \j_y-3 \left(7+128 \pi ^2\right) \j_y^2\right)\right)+3000 \pi  e_y^3 e_z-5 e_y^2 \j_z \left(\left(384 \pi ^2-839\right) \j_x+600 \pi  \j_y\right)-10 e_y e_z \left(120 \pi  \j_x^2+\left(576 \pi ^2-1225\right) \j_x \j_y+780 \pi  \j_y^2\right)+3 \j_z \left(67 \j_x^3-103 \j_x \j_y^2+40 \pi  \j_y^3\right)\right),\\
\frac{3}{128} \pi   \left(150 \pi  e_x^2 \left(15 e_z^2-11 \j_z^2\right)+15 e_x \left(7 e_y \left(15 e_z^2+61 \j_z^2\right)-20 e_z \j_z (20 \pi  \j_x+21 \j_y)\right)-25 e_z^2 \left(-30 \pi  \left(e_y^2+3 \j_y^2\right)+18 \pi  \j_x^2+203 \j_x \j_y\right)+3 \j_z^2 \left(10 \pi  \left(\j_y^2-45 e_y^2\right)+30 \pi  \j_x^2+21 \j_x \j_y\right)-100 e_y e_z \j_z (7 \j_x+36 \pi  \j_y)\right)\right\}.
\end{dmath*}

\label{lastpage}

\end{document}